\ifdefined\pdftexversion\pdfoutput=1\fi
\documentclass[11pt]{article}
\usepackage[letterpaper,margin=1in]{geometry}
\usepackage[T1]{fontenc}
\usepackage{lmodern}
\usepackage{microtype}
\usepackage{amsmath,amssymb}
\usepackage{booktabs,tabularx,array,multirow}
\usepackage{threeparttable}%
\providecommand\printbw{0}
\ifnum\printbw=1 \PassOptionsToPackage{gray}{xcolor}\fi
\usepackage[table]{xcolor}
\usepackage{graphicx}
\usepackage{tikz}
\usepackage{pgfplots}
\usepackage{pgfplotstable}
\pgfplotsset{compat=1.18}
\usepgfplotslibrary{groupplots}
\usetikzlibrary{positioning,patterns,arrows.meta,calc,decorations.pathreplacing,shapes.misc}
\usepackage{caption}
\usepackage{float}
\usepackage{enumitem}
\usepackage[numbers,sort&compress]{natbib}
\usepackage[colorlinks=true,linkcolor=blue!50!black,urlcolor=blue!50!black,citecolor=blue!50!black]{hyperref}
\ifnum\printbw=1 \hypersetup{hidelinks}\fi
\hypersetup{pdftitle={A small HER2 reader for routine H\&E slides: real-world deployment},
  pdfauthor={Thang Tran, Lan Dang}, pdfkeywords={HER2, breast cancer, computational pathology, H\&E whole-slide images, foundation models, knowledge distillation, JEPA, multiple-instance learning, PathBench, external validation, CPU inference, clinical deployment}}
\setlist{itemsep=1pt, topsep=3pt}

\newcommand\herprotoseeds{10}\newcommand\herprotoruns{20}

\providecommand\herprotoseeds{3}\providecommand\herprotoruns{6}
\providecommand\herset[2]{\expandafter\def\csname her:#1\endcsname{#2}}
\providecommand\her[1]{\ifcsname her:#1\endcsname\csname her:#1\endcsname\else\textbf{\textcolor{red}{??#1}}\fi}

\herset{rn50-yy}{0.701}
\herset{rn50-tt}{0.552}
\herset{rn50-ty}{0.630}
\herset{hibou-yy}{0.779}
\herset{hibou-tt}{0.644}
\herset{hibou-ty}{0.717}
\herset{phikon-yy}{0.876}
\herset{phikon-tt}{0.669}
\herset{phikon-ty}{0.813}
\herset{hoptaug-yy}{0.898}
\herset{hoptaug-tt}{0.669}
\herset{hoptaug-ty}{0.798}
\herset{gmac-phikon}{17.6}
\herset{cpu-phikon}{23.0}
\herset{par-hoptcard}{1.1B}
\herset{gmac-hoptcard}{304.2}
\herset{cpu-hoptcard}{312.3}
\herset{par-d16final}{22M}
\herset{gmac-d16final}{4.6}
\herset{cpu-d16final}{8.4}
\herset{gmac-hibou}{23.5}
\herset{cpux-phikon}{2.7}
\herset{cpux-hoptcard}{37}
\herset{ov-cohort-slides}{490}
\herset{ov-cohort-cases}{490}
\herset{ov-cohort-positive}{156}
\herset{ov-corpus-selected-slides}{9,752}
\herset{ov-corpus-selected-brca}{1,062}
\herset{ov-overlap-slides}{490}

\definecolor{hrCtl}{HTML}{7A7A7A}
\definecolor{hrS16}{HTML}{0072B2}
\definecolor{hrB16}{HTML}{009E73}
\definecolor{hrDst}{HTML}{E69F00}
\definecolor{hrRef}{HTML}{B5549A}
\definecolor{hrYY}{HTML}{0072B2}
\definecolor{hrTT}{HTML}{56B4E9}
\definecolor{hrTY}{HTML}{D55E00}
\definecolor{hrYT}{HTML}{E69F00}
\definecolor{hrBest}{HTML}{D7191C}%

\pgfdeclareplotmark{hrstar}{%
  \pgfpathmoveto{\pgfqpointpolar{90}{\pgfplotmarksize}}%
  \pgfpathlineto{\pgfqpointpolar{126}{0.42\pgfplotmarksize}}\pgfpathlineto{\pgfqpointpolar{162}{\pgfplotmarksize}}%
  \pgfpathlineto{\pgfqpointpolar{198}{0.42\pgfplotmarksize}}\pgfpathlineto{\pgfqpointpolar{234}{\pgfplotmarksize}}%
  \pgfpathlineto{\pgfqpointpolar{270}{0.42\pgfplotmarksize}}\pgfpathlineto{\pgfqpointpolar{306}{\pgfplotmarksize}}%
  \pgfpathlineto{\pgfqpointpolar{342}{0.42\pgfplotmarksize}}\pgfpathlineto{\pgfqpointpolar{18}{\pgfplotmarksize}}%
  \pgfpathlineto{\pgfqpointpolar{54}{0.42\pgfplotmarksize}}\pgfpathclose\pgfusepathqfillstroke}

\pgfplotsset{
  hr groups/.style={scatter, scatter src=explicit symbolic, scatter/classes={
      ctl={mark=*, hrCtl}, s16={mark=*, hrS16}, b16={mark=square*, hrB16},
      dst={mark=diamond*, hrDst, mark size=2.6pt}, dstb={mark=pentagon*, hrB16, mark size=2.8pt},
      ref={mark=triangle*, hrRef, mark size=2.6pt},
      best={mark=hrstar, draw=black, fill=hrBest, mark size=4.2pt}}},
  hr lead/.style={scatter, scatter src=explicit symbolic, scatter/classes={
      best={mark=hrstar, draw=black, fill=hrBest, mark size=4.6pt}, ours={mark=diamond*, hrS16, mark size=2.8pt},
      ref={mark=triangle*, hrRef, mark size=2.8pt}, ctl={mark=*, hrCtl}}},
  hr panel/.style={tick label style={font=\scriptsize}, label style={font=\small}, title style={font=\small},
    grid style={black!8}, legend style={font=\scriptsize, draw=black!25, fill=white, fill opacity=0.9, text opacity=1}},
}

\newcommand\herprotocolfigure{%
\begin{tikzpicture}[font=\footnotesize, >=stealth, x=1mm, y=1mm,
  hp/.style={draw=black!55, rounded corners=2pt, align=center, inner sep=2.5pt, fill=##1}, hp/.default=white,
  ha/.style={->, draw=black!60, thick}, hl/.style={font=\scriptsize\itshape, text=black!65, align=center}]
\node[hp=hrS16!12, text width=19mm] (enc) at (0,0) {\textbf{encoder}\\under test\\[1pt]\emph{frozen}};
\node[hp=black!4, text width=29mm] (by) at (34,15) {\textbf{Yale-HER2}, 192 slides\\400 tiles per slide};
\node[hp=black!4, text width=29mm] (bt) at (34,-15) {\textbf{TCGA-BRCA}, 490 slides\\400 tiles per slide};
\node[hp=hrDst!18, text width=29mm] (cy) at (74,15) {\textbf{CLAM on Yale}\\5-fold CV, \herprotoseeds\ seeds};
\node[hp=hrDst!18, text width=29mm] (ct) at (74,-15) {\textbf{CLAM on TCGA}\\5-fold CV, \herprotoseeds\ seeds};
\node[hp=hrYY!14, text width=40mm] (yy) at (121,22) {\textbf{Yale$\to$Yale}: pooled\\out-of-fold AUC};
\node[hp=hrYT!18, text width=40mm] (yt) at (121,8) {\textbf{Yale$\to$TCGA}: mean of the\\5 fold models, on all of TCGA};
\node[hp=hrTT!18, text width=40mm] (tt) at (121,-8) {\textbf{TCGA$\to$TCGA}: pooled\\out-of-fold AUC};
\node[hp=hrTY!14, text width=40mm] (ty) at (121,-22) {\textbf{TCGA$\to$Yale}: mean of the\\5 fold models, on all of Yale};
\draw[ha] (enc.east) -- (12,0) |- (by.west);
\draw[ha] (enc.east) -- (12,0) |- (bt.west);
\draw[ha] (by) -- (cy); \draw[ha] (bt) -- (ct);
\draw[ha] (cy.east) -- (95,15) |- (yy.west); \draw[ha] (cy.east) -- (95,15) |- (yt.west);
\draw[ha] (ct.east) -- (95,-15) |- (tt.west); \draw[ha] (ct.east) -- (95,-15) |- (ty.west);
\draw[->, draw=black!40, dashed] (bt.east) -- (cy.south west);
\draw[->, draw=black!40, dashed] (by.east) -- (ct.north west);
\node[hl, fill=white, inner sep=1pt] at (54,0) {each cohort is the other's\\external test};
\node[hl, text width=34mm, anchor=north] at (34,-23) {feature extraction: the slow step\\(5--10\,h for TCGA, read over the network)};
\node[hl, text width=30mm, anchor=north] at (74,-23) {\herprotoruns\ runs of about\\1 minute each};
\node[hl, text width=40mm, anchor=north] at (121,-29) {four numbers per encoder,\\mean (SD) over \herprotoseeds\ seeds};
\end{tikzpicture}}

\providecommand\mset[2]{\expandafter\def\csname m5:#1\endcsname{#2}}
\mset{phikon-yy}{0.872}
\mset{phikon-tt}{0.679}
\mset{phikon-ty}{0.811}
\mset{phikon-yt}{0.662}
\mset{hopt-ty}{0.805}
\mset{sfinal-yy}{0.761}
\mset{sfinal-tt}{0.589}
\mset{sfinal-ty}{0.589}
\mset{sfinal-yt}{0.520}
\mset{m5s25-yy}{0.858}
\mset{m5s25-tt}{0.705}
\mset{m5s25-ty}{0.762}
\mset{m5s25-yt}{0.625}
\mset{m5b50-ty}{0.783}
\mset{jj25s3-yy}{0.844}
\mset{jj25s3-tt}{0.690}
\mset{jj25s3-ty}{0.812}
\mset{jj25s3-yt}{0.650}
\mset{jj25s4-ty}{0.812}
\mset{jj25s5-ty}{0.810}
\mset{wisea50-yy}{0.859}
\mset{wisea50-ty}{0.797}
\mset{j100s-yy}{0.870}
\mset{j100s-ty}{0.776}
\mset{pd-k1b-m5b25-ty}{$-$0.020}
\mset{pd-k1s-m5s25-yt}{$-$0.035}
\mset{pd-m5b75-m5b50-ty}{$-$0.043}
\mset{pd-m6ec-m6ef-yy}{+0.016}
\mset{pd-m6ec-m6ef-yt}{+0.013}
\mset{pd-v5s-v4s-ty}{$-$0.024}
\mset{pd-jms3-m6ds-yy}{$-$0.016}
\mset{pd-jms3-m6ds-ty}{+0.027}
\mset{pd-jlr3-m6ds-ty}{+0.014}
\mset{pd-jsh3-m6ds-yy}{$-$0.039}
\mset{pd-j25s3-m5s25-ty}{+0.036}
\mset{pd-j25s4-m5s25-ty}{+0.030}
\mset{pd-jj25s3-j25s3-ty}{+0.014}
\mset{pd-jj25s3-j25s3-yt}{+0.007}
\mset{pd-jj25s3-m5s25-yy}{$-$0.014}
\mset{pd-jj25s3-m5s25-tt}{$-$0.014}
\mset{pd-jjj25s3-jj25s3-yy}{$-$0.010}
\mset{pd-jjj25s3-jj25s3-ty}{$-$0.006}
\mset{pd-jaug25-k4s-ty}{$-$0.006}
\mset{pd-j25aug-j25s3-ty}{$-$0.002}
\mset{pd-j25blk-j25s3-ty}{$-$0.008}
\mset{pd-j25lr2-j25s3-ty}{$-$0.017}
\mset{pd-j100s-m5s-yy}{+0.030}
\mset{pd-rda-jms3-ty}{$-$0.049}
\mset{pd-jj25s3-phikon-tt}{+0.011}
\mset{pd-jj25s3-phikon-tt-k}{8/10}
\mset{pd-jj25s3-phikon-tt-p}{0.042}
\mset{pd-jj25s3-phikon-ty}{+0.000}
\mset{pd-jj25s3-phikon-ty-k}{5/10}
\mset{pd-jj100s3-j100s-yy}{$-$0.027}
\mset{pd-jj100s3-j100s-ty}{$-$0.000}
\mset{pd-j2512k-jj25s3-ty}{$-$0.024}
\mset{pd-jj25s3-hopt-ty-k}{7/10}
\mset{pd-mixa75-jj25s3-yy}{+0.003}
\mset{pd-mixa75-jj25s3-ty}{$-$0.018}
\mset{pd-mixa50-jj25s3-yy}{+0.002}
\mset{pd-mixa50-jj25s3-ty}{$-$0.017}
\mset{pd-xtf-jj25s3-tt}{$-$0.026}
\mset{pd-xtf-jj25s3-ty}{$-$0.009}
\mset{pd-xth-jj25s3-tt}{$-$0.038}
\mset{pd-xth-jj25s3-ty}{$-$0.019}
\mset{pd-wisea75-jj25s3-yy}{+0.011}
\mset{pd-wisea75-jj25s3-ty}{$-$0.011}
\mset{pd-wisea50-jj25s3-yy}{+0.015}
\mset{pd-wisea50-jj25s3-ty}{$-$0.015}

\providecommand\mm[1]{\ifcsname m5:#1\endcsname\csname m5:#1\endcsname\else\textbf{\textcolor{red}{??#1}}\fi}
\providecommand\pset[2]{\expandafter\def\csname pp:#1\endcsname{#2}}
\pset{r1-yy}{0.850}
\pset{r1-tt}{0.688}
\pset{r1-ty}{0.795}
\pset{r1-ty-spread}{0.006}
\pset{r1-yt}{0.642}
\pset{rank-n}{7}
\pset{pb-tcga-ours}{0.689}
\pset{pb-yale-ours}{0.897}
\pset{pb-n}{7}
\pset{t4-her2low-ours}{0.649}
\pset{t4-her2low-rank}{4}
\pset{t4-her2low-phikon}{0.741}
\pset{t4-her2low-hopt}{0.692}
\pset{t4-fish-min}{0.47}
\pset{t4-fish-max}{0.55}
\pset{t4-fishx-ours}{0.682}
\pset{t4-fishx-phikon}{0.619}
\pset{t4-fishx-hopt}{0.706}
\pset{t4-fisho-ours}{0.669}
\pset{t4-lowx-min}{0.44}
\pset{t4-lowo-max}{0.57}
\pset{t4-fish-n}{143}
\pset{t4-fish-pos}{44}
\pset{t4-fish-agree}{131}
\pset{t4-fish-both}{143}
\pset{t4-n}{7}
\pset{t4s2-her2low}{0.683}
\pset{rd-thr}{0.27}
\pset{rd-n-yale}{192}
\pset{rd-n-tcga}{490}
\pset{rd-yale-sens}{0.72}
\pset{rd-yale-sens-ci}{0.62--0.80}
\pset{rd-yale-spec}{0.79}
\pset{rd-yale-spec-ci}{0.70--0.86}
\pset{rd-yale-ppv}{0.76}
\pset{rd-yale-npv}{0.75}
\pset{rd-yale05-sens}{0.14}
\pset{rd-yale05-sens-ci}{0.08--0.22}
\pset{rd-yale05-spec}{0.97}
\pset{rd-yale05-spec-ci}{0.91--0.99}
\pset{rd-yale05-ppv}{0.81}
\pset{rd-yale05-npv}{0.55}
\pset{rd-tcga-sens}{0.58}
\pset{rd-tcga-spec}{0.75}
\pset{rd-tcga-ppv}{0.51}
\pset{rd-tcga-npv}{0.79}
\pset{rd-auc}{0.815}
\pset{rd-auc-ci}{0.75--0.88}

\providecommand\pp[1]{\ifcsname pp:#1\endcsname\csname pp:#1\endcsname\else\textbf{\textcolor{red}{??#1}}\fi}
\providecommand\probeset[2]{\expandafter\def\csname probe:#1\endcsname{#2}}

\definecolor{pbBlue}{HTML}{0072B2}     %
\definecolor{pbGreen}{HTML}{009E73}    %
\definecolor{pbPurple}{HTML}{B5549A}   %
\definecolor{pbVermil}{HTML}{D55E00}   %
\definecolor{pbSky}{HTML}{3A9AD9}      %
\definecolor{pbGrey}{HTML}{7A7A7A}     %
\definecolor{pbBlack}{HTML}{303030}    %

\pgfplotsset{
  pb vits16/.style={color=pbBlue, mark=*},
  pb vitb16/.style={color=pbGreen, mark=square*},
  pb st-midnight/.style={color=pbPurple, mark=diamond*},
  pb st-hopt/.style={color=pbVermil, mark=diamond*},
  pb midnight/.style={color=pbPurple, mark=*},
  pb hopt/.style={color=pbVermil, mark=*},
  pb phikon/.style={color=pbSky, mark=*},
  pb dinov2/.style={color=pbGrey, mark=*},
  pb control/.style={color=pbBlack, mark=*},
  pb curve/.style={very thick, mark size=1.3pt, mark options={solid, fill=white, line width=0.6pt}},
  pb band/.style={draw=none, fill opacity=0.16, forget plot},
  pb ref/.style={line width=0.55pt, dashed, dash pattern=on 3pt off 1.5pt, forget plot},
  pb panel/.style={scale only axis, height=56mm, ymin=0.05, ymax=0.80, ytick={0.1,0.2,...,0.8},
    yticklabel style={/pgf/number format/fixed, /pgf/number format/precision=1},
    tick label style={font=\scriptsize}, label style={font=\small}, title style={font=\small, yshift=-1mm},
    ymajorgrids, grid style={black!8}, clip mode=individual,
    legend cell align=left, legend style={font=\scriptsize, draw=black!25, fill=white, fill opacity=0.9,
      text opacity=1, row sep=-1pt, at={(0.98,0.03)}, anchor=south east}},
}

\tikzset{
  arB/.style={draw=black!55, rounded corners=2pt, align=center, inner sep=2.5pt, font=\scriptsize, fill=#1},
  arB/.default=white,
  arA/.style={->, >=stealth, draw=black!65, thick},
  arT/.style={font=\scriptsize, text=black!75, align=center},
  arN/.style={font=\scriptsize\itshape, text=black!60, align=center},
}

\newcommand\archreaderfigure{%
\begin{tikzpicture}[x=1mm, y=1mm]
\node[arB=black!4, text width=30mm, minimum height=16mm] (wsi) at (0,0) {\textbf{Whole-slide image}\\H\&E, read from the\\image archive};
\node[arB=black!4, text width=34mm, minimum height=16mm] (sel) at (50,0) {\textbf{Tile selection}\\tissue mask; up to 400 tiles\\of 256\,px at 0.5\,\textmu m/px\\(128\,\textmu m), as in training};
\node[arB=hrS16!14, text width=34mm, minimum height=16mm] (bb) at (110,0) {\textbf{Backbone}: ViT-S/16\\21.7M parameters, frozen;\\runs once per tile};
\node[arB=black!4, text width=30mm, minimum height=16mm] (bag) at (160,0) {\textbf{Bag of features}\\400 $\times$ 384:\\one vector per tile};
\node[arB=hrDst!20, text width=34mm, minimum height=16mm] (clam) at (160,-40) {\textbf{CLAM-SB} $\times$ 5\\one per CV fold;\\0.59M parameters each;\\classifier built in};
\node[arB=black!4, text width=34mm, minimum height=16mm] (ens) at (104,-40) {\textbf{Ensemble}\\mean of the 5 P(HER2+);\\mean of the 5 attentions};
\node[arB=black!4, text width=34mm, minimum height=16mm] (op) at (50,-40) {\textbf{Operating point}\\per laboratory:\\threshold $\to$ proposed call};
\node[arB=hrB16!16, text width=30mm, minimum height=16mm] (out) at (0,-40) {\textbf{Proposal}\\P(HER2+), call,\\attention heatmap,\\model-pack identity};
\draw[arA] (wsi) -- (sel);
\draw[arA] (sel) -- node[arT, above=1mm] {400$\times$3$\times$224$^2$} (bb);
\draw[arA] (bb) -- node[arT, above=1mm] {CLS} (bag);
\draw[arA] (bag) -- node[arT, right=1mm] {400$\times$384} (clam);
\draw[arA] (clam) -- node[arT, above=1mm] {5 P, 5 maps} (ens);
\draw[arA] (ens) -- (op);
\draw[arA] (op) -- (out);
\node[arB=white, text width=30mm, dashed] (path) at (0,-65) {\textbf{pathologist}\\accepts or declines};
\draw[arA] (out) -- (path);
\node[arN, anchor=north] at (110,-10.5) {learned without labels:\\\archbackbonetext};
\node[arN, anchor=north] at (50,-10.5) {a fixed rule, identical in\\training and deployment};
\node[arN, anchor=north] at (160,-50.5) {learned from HER2 status labels:\\TCGA-BRCA, 5-fold CV};
\node[arN, anchor=north] at (50,-50.5) {set on validation slides,\\then per laboratory};
\node[arN, anchor=north] at (104,-50.5) {a fixed rule};
\end{tikzpicture}}

\definecolor{cdata}{HTML}{2E86AB}\definecolor{cmodel}{HTML}{A23B72}\definecolor{cgpu}{HTML}{F18F01}
\definecolor{cguard}{HTML}{C73E1D}\definecolor{cok}{HTML}{3B8E4F}\definecolor{cgrey}{HTML}{6C757D}
\tikzset{
  box/.style={draw, rounded corners=2pt, align=center, font=\small, minimum height=7mm, inner sep=3pt},
  dbox/.style={box, fill=cdata!12, draw=cdata!80!black},
  mbox/.style={box, fill=cmodel!10, draw=cmodel!80!black},
  gbox/.style={box, fill=cgpu!14, draw=cgpu!80!black},
  rbox/.style={box, fill=cguard!10, draw=cguard!80!black},
  okbox/.style={box, fill=cok!12, draw=cok!80!black},
  arr/.style={-{Stealth[length=2.2mm]}, thick},
  darr/.style={-{Stealth[length=2.2mm]}, thick, dashed},
  lbl/.style={font=\scriptsize, align=center, text=black!75},
}
\newcommand\booksizes{\def\scriptsize{\fontsize{7}{8.5}\selectfont}\def\footnotesize{\fontsize{8}{9.5}\selectfont}%
  \def\small{\fontsize{9}{11}\selectfont}}

\newcommand{\enc}[1]{\textsf{#1}}   %

\newcommand{\YY}{Yale$\to$Yale}
\newcommand{\TT}{TCGA$\to$TCGA}
\newcommand{\TY}{TCGA$\to$Yale}
\newcommand{\YT}{Yale$\to$TCGA}

\title{{\Large\textbf{A small HER2 reader for routine H\&E slides: real-world deployment}}}
\author{%
  Thang Tran\thanks{Corresponding author: \texttt{thang.tran@cloudkites.com}}\\
  \small CloudKites AI Lab\\
  \small New South Wales, Australia\\
  \small\texttt{thang.tran@cloudkites.com}
  \and
  Lan Dang\\
  \small Monash Business School, Monash University\\
  \small Victoria, Australia\\
  \small\texttt{LanHong.Dang@monash.edu}
}
\date{October 2026}

\begin{document}
\maketitle

\begin{abstract}
\noindent HER2 status decides who receives HER2-targeted therapy. Pathologists score HER2 protein by
immunohistochemistry (IHC); equivocal IHC~2+ tumours, their hardest calls, wait for in situ hybridisation (ISH). We
built a HER2 reader for routine H\&E slides, owned end to end, as a baseline for pathologists. A 22M-parameter ViT-S/16
tile encoder was trained on public TCGA slides by self-supervision, distillation from H-optimus-0 and two rounds of JEPA
refinement; five cross-validated attention-based slide classifiers average their probabilities; and an operating point
is fixed on training data. Compared under one paired ten-seed protocol, readers on our encoder rank first of
\pp{rank-n} on external validation from TCGA-BRCA to Yale-HER2 (AUC \mm{jj25s3-ty}), level with readers on encoders 4
and 50 times larger. We also rebuilt PathBench's encoder benchmark on public cohorts; our replica reproduces its published
order of reference encoders, and under it our encoder ranks first of \pp{pb-n} on TCGA-BRCA (\pp{pb-tcga-ours}) and sits
level with H-optimus-0 and Phikon on Yale-HER2 (\pp{pb-yale-ours}). On three external hospitals our deployed reader
reaches AUC 0.82 on Yale-HER2 (about 0.81 with training slides balanced for tissue size), 0.75 on HEROHE and 0.63 on
BCNB biopsies, and among HEROHE's 85 IHC~2+ cases it ranks ISH-amplified above non-amplified tumours at 0.76, best among
our TCGA-trained readers. Per-laboratory score normalisation and a laboratory's own slides improve it
further on a new site. Without a GPU our reader needs 5--6\,s per slide. It serves
as a research-use pre-read in a pathology platform, where it proposes a probability and a call beside IHC and helps
prioritise ISH for equivocal cases. Weights, per-slide predictions and the data behind every figure are public at
\url{https://huggingface.co/cloudkites/her2-reader}.
\end{abstract}
\begin{quote}\small\vspace{-1.2ex}\noindent\textbf{Keywords:} HER2; breast cancer; computational pathology; H\&E whole-slide images; foundation models; knowledge distillation; JEPA; multiple-instance learning; PathBench; external validation; CPU inference; clinical deployment.\end{quote}

\section{Introduction}\label{sec:intro}

HER2 status decides who receives HER2-targeted therapy, and it can be read at three levels of one causal chain
(Figure~\ref{fig:proto-levels}). ISH counts copies of the \emph{ERBB2} gene. IHC stains HER2 receptor protein on tumour
cell membranes and is scored 0, 1+, 2+ or 3+ by staining intensity and completeness
(Figure~\ref{fig:proto-ihc})~\citep{wolff2018,wolff2023}. Routine H\&E shows only tissue that amplification has shaped.
IHC~3+ counts as positive, 0 or 1+ as negative. An IHC~2+ tumour is \emph{equivocal} and goes to ISH, which costs more,
takes days and is not available everywhere (Table~\ref{tab:proto-tests}). Since trastuzumab deruxtecan proved effective
in HER2-low disease (IHC~1+, or 2+ without amplification)~\citep{modi2022}, both a 2+ call and a call between 0 and 1+
change treatment, and pathologists agree poorly on both~\citep{fernandez2022}.

{\booksizes
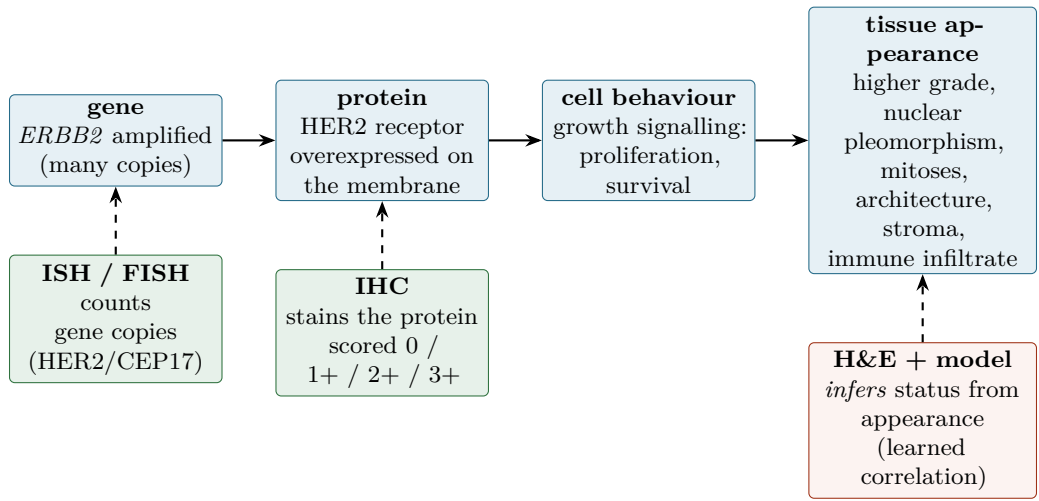
\begin{figure}[!htbp]
\centering
\begin{tikzpicture}[font=\scriptsize, node distance=5mm and 7mm]
\node[dbox, text width=26mm] (g) {\textbf{gene}\\\emph{ERBB2} amplified\\(many copies)};
\node[dbox, right=of g, text width=26mm] (p) {\textbf{protein}\\HER2 receptor\\overexpressed on\\the membrane};
\node[dbox, right=of p, text width=26mm] (b) {\textbf{cell behaviour}\\growth signalling:\\proliferation,\\survival};
\node[dbox, right=of b, text width=28mm] (m) {\textbf{tissue appearance}\\higher grade, nuclear\\pleomorphism, mitoses,\\architecture, stroma,\\immune infiltrate};
\draw[arr] (g) -- (p); \draw[arr] (p) -- (b); \draw[arr] (b) -- (m);
\node[okbox, below=9mm of g, text width=26mm] (ish) {\textbf{ISH / FISH}\\counts gene copies\\(HER2/CEP17)};
\node[okbox, below=9mm of p, text width=26mm] (ihc) {\textbf{IHC}\\stains the protein\\scored 0 / 1+ / 2+ / 3+};
\node[rbox, fill=cguard!6, below=9mm of m, text width=28mm] (he) {\textbf{H\&E + model}\\\emph{infers} status from\\appearance (learned\\correlation)};
\draw[darr] (ish) -- (g); \draw[darr] (ihc) -- (p); \draw[darr] (he) -- (m);
\node[lbl, anchor=north west, text width=120mm, align=left] at ($(ish.south west |- he.south)+(0,-3mm)$) {H\&E sees no HER2 molecule: only the downstream consequences that the gene and the protein leave in the tissue.};
\end{tikzpicture}
\caption{Three tests read three different links of one causal chain. ISH and IHC measure HER2 directly; an
H\&E model reads the chain's end and infers the start.}
\label{fig:proto-levels}
\end{figure}}
{\booksizes\definecolor{ihcnone}{HTML}{F4E3A1}
\definecolor{ihcfaint}{HTML}{E0A93B}
\definecolor{ihcmid}{HTML}{9C5A1E}
\definecolor{ihcdark}{HTML}{4A2508}
\definecolor{ihccyto}{HTML}{FBF3DC}
\definecolor{ihcnuc}{HTML}{4F6FA8}
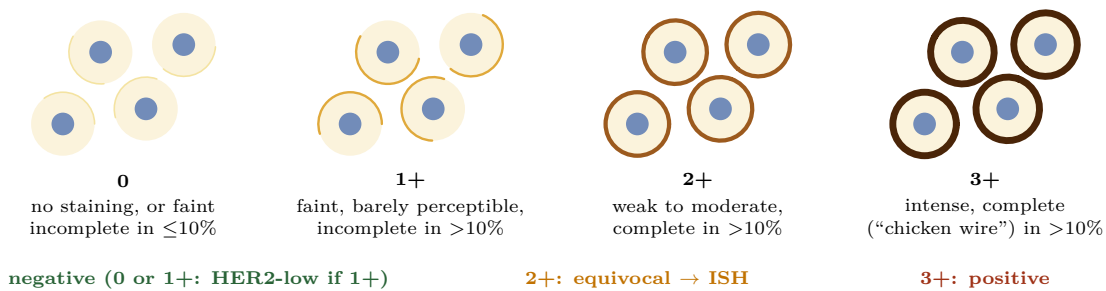
\begin{figure}[!htbp]
\centering
\begin{tikzpicture}[font=\scriptsize]
\foreach \i/\lbl/\ring/\lw/\frac/\txt in {
    0/0/ihcnone/0.6pt/0.35/{no staining, or faint\\incomplete in $\le$10\%},
    1/1+/ihcfaint/0.9pt/0.55/{faint, barely perceptible,\\incomplete in $>$10\%},
    2/2+/ihcmid/1.6pt/1.0/{weak to moderate,\\complete in $>$10\%},
    3/3+/ihcdark/2.6pt/1.0/{intense, complete\\(``chicken wire'') in $>$10\%}} {
  \begin{scope}[xshift=\i*38mm]
    \foreach \x/\y/\rot in {0/0/0, 1.1/0.2/70, 0.5/0.95/150, 1.6/1.05/230} {
      \fill[ihccyto] (\x,\y) circle (0.42);
      \fill[ihcnuc!80] (\x,\y) circle (0.15);
      \pgfmathsetmacro{\arc}{360*\frac}
      \draw[draw=\ring, line width=\lw, line cap=round] ([shift={(\rot:0.42)}]\x,\y) arc[start angle=\rot, delta angle=\arc, radius=0.42];
    }
    \node at (0.8,-0.75) {\textbf{\lbl}};
    \node[align=center] at (0.8,-1.25) {\txt};
  \end{scope}
}
\node[align=center, font=\scriptsize\bfseries, text=cok!70!black] at (1.8,-2.05) {negative (0 or 1+: HER2-low if 1+)};
\node[align=center, font=\scriptsize\bfseries, text=cgpu!80!black] at (4*1.9,-2.05) {2+: \textbf{equivocal} $\to$ ISH};
\node[align=center, font=\scriptsize\bfseries, text=cguard!80!black] at (11.4+0.8,-2.05) {3+: positive};
\end{tikzpicture}
\caption{\emph{Schematic.} HER2 immunohistochemistry: the brown (DAB) ring on the membrane of the invasive tumour cells, blue
haematoxylin nuclei. ASCO/CAP grade it by intensity and completeness over the proportion of tumour
cells~\cite{wolff2018,wolff2023}: 0 no or faint incomplete staining in $\le$10\%; 1+ faint, incomplete in $>$10\%; 2+
weak to moderate complete staining in $>$10\% (equivocal, resolved by ISH); 3+ intense, complete staining in $>$10\%. 1+,
or 2+ with negative ISH, is \emph{HER2-low}, now treatable with trastuzumab deruxtecan~\cite{modi2022}.}
\label{fig:proto-ihc}
\end{figure}
}

\paragraph{Where an H\&E reader helps.} A \emph{HER2 reader} predicts HER2 status from an H\&E slide, which every case
already has. A pathology foundation model encodes each tissue tile, and an attention-based classifier pools tile
features into one slide-level probability~\citep{lu2021,lv2026}. Such a reader cannot see protein, so it replaces neither
IHC nor ISH, and it cannot grade HER2-low. It can give a first signal before any IHC section is cut, flag a case whose
confident prediction disagrees with its IHC score, and rank equivocal 2+ tumours so that ISH goes first to those most
likely amplified. That last use matters most, because 2+ is where pathologists find a decision hardest; it is worth
having only if a reader holds up on hospitals it has never seen.

{\booksizes\begin{table}[!htbp]
\centering\small
\caption{The three tests compared.}\label{tab:proto-tests}
\begin{tabularx}{\textwidth}{@{}lXXX@{}}
\toprule
 & \textbf{H\&E + model} & \textbf{IHC} & \textbf{ISH} \\
\midrule
measures & tissue appearance; \emph{infers} HER2 & HER2 protein on the membrane & \emph{ERBB2} gene copies \\
result & probability, heatmap & 0 / 1+ / 2+ / 3+ & ratio; amplified or not \\
when available & first: every case has an H\&E & after a dedicated stained section & days later; only for 2+ (or on request) \\
relative cost (design doc) & compute only & about \$30--118 & about \$160--320 \\
weak points & indirect; learned correlations; confounded by grade and site & 2+ equivocal; poor agreement at 0 vs 1+; sensitive to fixation & cost, time, availability \\
\bottomrule
\end{tabularx}
\end{table}
}

\paragraph{Which encoders read HER2 best?} Published benchmarks keep naming a small group of models. H-optimus-0, a
1.1B-parameter ViT-g/14 trained on over 500k slides~\citep{hoptimus2024}, comes first on average among six encoders across
six HER2 cohorts in one study, where a reader built on it and trained on TCGA and CPTAC reaches AUC 0.91 on
Yale-HER2~\citep{athena2025}. It also comes first on breast HER2 in PathBench~\citep{pathbench2025}, and its successor
leads on Yale-HER2 and on HEROHE's IHC~2+ cases in a third study~\citep{lv2026}. Phikon, an 86M-parameter ViT-B/16 trained
with iBOT on TCGA~\citep{filiot2023}, makes a natural yardstick for a compact reader.

\paragraph{Why a small, owned reader?} Most HER2 models run on GPUs. A reader 4 to 50 times smaller costs less per slide
and can run where no GPU exists, such as an ordinary CPU server in a hospital's container cluster. On a 12-core CPU a
ViT-S/16 (22M parameters) encodes a slide in \her{cpu-d16final}\,s, against \her{cpu-phikon}\,s for Phikon's ViT-B/16 and
about 5\,min for H-optimus-0 (\S\ref{sec:res-cpu}). A product must also own its model: most released pathology encoders,
and CLAM's reference code, carry licences that a commercial reader cannot inherit. Yet small encoders trained without a
foundation model transfer poorly between cohorts, so being small is not enough.

\paragraph{What we built and tested.} We built a complete HER2 reader on a ViT-S/16, using only our own code and public
data (Figure~\ref{fig:arch-reader}). Its tile encoder was trained on TCGA slides in three stages (Figure~\ref{fig:stages}):
self-supervised pretraining, distillation of H-optimus-0's tile features, and JEPA refinement, in which our distilled
encoder learns to predict its own features of hidden patches~\citep{assran2023,genbio2026}. CLAM slide classifiers trained
in five folds average their probabilities, and an operating point fixed on training data turns that average into a call.
We compared readers that differ only in their encoder under two independent protocols: one paired protocol run in both
directions between two cohorts, and a local replica of PathBench's benchmark that runs on public data. We then judged our
reader as a whole on three hospitals it never trained on (Yale-HER2, HEROHE and BCNB), including HEROHE's 85 equivocal
IHC~2+ tumours, and compared it with the best published reader under that reader's own protocol. It runs as a research-use pre-read in a
pathology platform, next to that platform's ER, PR and Ki-67 results (\S\ref{sec:method-deploy}).

\paragraph{Contributions.}
\begin{enumerate}[leftmargin=*,itemsep=1pt]
\item An owned HER2 reader of 22M encoder parameters and five 0.6M-parameter slide classifiers that ranks first of
  \pp{rank-n} on external validation (TCGA-BRCA$\to$Yale-HER2, AUC \mm{jj25s3-ty}) among readers that differ only in their
  encoder, level with Phikon at a quarter of its size and ahead of H-optimus-0 at a fiftieth (\S\ref{sec:res-rank}). Its
  encoder comes from a three-stage recipe whose stages we measure and replicate (\S\ref{sec:res-stages}).
\item That reader judged on three external hospitals and on equivocal IHC~2+ tumours: AUC 0.82 on Yale-HER2, 0.75 on
  HEROHE, 0.63 on BCNB, and 0.76 for ISH amplification among HEROHE's IHC~2+ cases, best among our TCGA-trained
  readers (\S\ref{sec:res-hospitals}).
\item A local PathBench protocol that ranks encoders for HER2 on public cohorts: a fresh attention-MIL classifier per
  task, encoder and seed, shared case-stratified 7:1:2 splits over ten paired seeds, and exact paired tests. It reproduces
  PathBench's published order of reference encoders on breast HER2 on both cohorts, so PathBench-style rankings can be
  rerun and extended without its private data. Under it our encoder ranks first of \pp{pb-n} on TCGA-BRCA (AUC
  \pp{pb-tcga-ours}) and sits level with H-optimus-0 and Phikon on Yale-HER2 (\pp{pb-yale-ours}; \S\ref{sec:res-pb},
  Supplementary~\ref{app:pb}).
\item A deployment: one verified model pack serving a CPU pre-read in a pathology platform at 5--6\,s per slide, two
  per-laboratory levers (site score normalisation and a site's own slides) that improve readers on a new site, and a
  review that sets every reader we built beside published state of the art and keeps this one as pathologists' baseline
  (\S\ref{sec:method-deploy}, \S\ref{sec:res-bias}, \S\ref{sec:res-review}). Weights and per-slide predictions are public.
\end{enumerate}

\section{Related work}\label{sec:related}

\paragraph{Self-supervised objectives.} DINO matches a student to an exponential-moving-average teacher's prototype
assignment across augmented views~\citep{caron2021,caron2020}; iBOT adds a masked-patch target~\citep{zhou2022}; DINOv2
combines both with a KoLeo regulariser at scale~\citep{sablayrolles2019,oquab2024}. Later work replaced KoLeo with a
kernel-density uniformity term~\citep{wang2020,karasikov2025} and added register tokens~\citep{darcet2024}. Masked
autoencoders~\citep{he2022} predict masked content in pixel space.

\paragraph{Joint-embedding predictive refinement.} I-JEPA predicts a target encoder's embeddings of masked image
blocks from a visible context~\citep{assran2023}, with no pixel reconstruction and no hand-crafted view augmentation. In
pathology, a DINO stage followed by a JEPA stage on its frozen stage-one teacher has been reported to make features hold
up better across sites and scanners~\citep{genbio2026}. We apply a JEPA stage to a \emph{distilled} student, with that
student as its own frozen target. We know of no evaluation of this pairing, feature distillation from a foundation model
followed by self-predictive refinement, for cross-cohort HER2 prediction.

\paragraph{Pathology foundation encoders.} Phikon applies iBOT to a ViT-B/16 on TCGA~\citep{filiot2023}, and Phikon-v2 is a ViT-L trained with DINOv2 on
460M tiles from over 100 public cohorts~\citep{filiot2024phikonv2}. UNI~\citep{chen2024uni}, Virchow~\citep{vorontsov2024virchow},
Virchow2~\citep{zimmermann2024virchow2} and Hibou~\citep{nechaev2024hibou} train DINOv2-style ViTs on institutional or
proprietary collections. H-optimus-0 is a 1.1B-parameter ViT-g/14 trained on a proprietary collection of over 500k
slides~\citep{hoptimus2024}; Midnight-12k is a ViT-g/14 trained on 12k TCGA slides~\citep{karasikov2025}. One study argues that slide diversity matters more than patch count~\citep{athena2025}. We have used our numbat
framework before to train and verify a detector, a language-model fine-tune and an endoscopy foundation
model~\citep{tran2026numbat,tran2026crossstack,tran2026woma}.

\paragraph{Distillation into compact encoders.} Knowledge distillation~\citep{hinton2015,romero2015} has been used to
cut inference cost of pathology encoders: H0-mini distils H-optimus-0~\citep{filiot2025}, DistillPath distils each of four
released encoders into a pretrained ViT-S/16~\citep{distillpath2026}, Pathryoshka combines multi-teacher distillation
with nested (Matryoshka) embeddings~\citep{pathryoshka2025},
and GPFM uses unified multi-teacher distillation~\citep{gpfm2025}. A student inherits its teacher's data provenance,
which we report beside every student.

\paragraph{Stain variation and site effects.} Perturbing stain concentrations after colour
deconvolution~\citep{ruifrok2001} was among augmentations that helped generalisation most in one
comparison~\citep{tellez2019}, and site-specific signatures inflate within-TCGA accuracy~\citep{howard2021}.

\paragraph{Slide aggregation and HER2 benchmarks.} Attention-based multiple-instance learning (ABMIL)~\citep{ilse2018}
pools tile embeddings into a slide prediction; CLAM adds instance-level clustering~\citep{lu2021} and is used in recent HER2
benchmarks~\citep{lv2026}; state-space alternatives exist~\citep{mambamilplus2025,2dmamba2024}. Lv et al.\ report
H-optimus-1 best on Yale-HER2 under cross-validation (0.900) and on their one external cohort~\citep{lv2026}; Bosch et
al.\ report H-optimus-0 highest on average, though not on every cohort, among six encoders across six HER2
cohorts~\citep{athena2025}; PathBench, on private cohorts held out of all pretraining, ranks H-optimus-0 first on
breast HER2 (0.845)~\citep{pathbench2025}. None of them reports transfer in both directions between cohorts.

\paragraph{HER2 from H\&E on external hospitals.} The HEROHE challenge released 509 HER2-annotated breast cancer slides,
rich in equivocal IHC~2+ cases (85 of 150 in its test split), and found H\&E-only prediction feasible but
hard~\citep{conde2022herohe}. With foundation encoders and CLAM, Virchow2 reaches AUC 0.86 on HEROHE and H-optimus-1
0.82 on its IHC~2+ cases~\citep{lv2026}. BCNB's 1,058 core-needle biopsies, collected for lymph-node
prediction~\citep{xu2021bcnb}, make a harder external test; there the best published reader reaches
0.72~\citep{athena2025}. That reader, H-optimus-0 with
attention MIL over every tissue tile at 1.0\,\textmu m/px, trained on TCGA-BRCA and CPTAC-BRCA~\citep{krug2020cptac} and
ensembled over 40 models, reaches 0.91 on Yale-HER2~\citep{athena2025}; we replicate its protocol with our encoder
(\S\ref{sec:res-protocol}).

\section{Data}\label{sec:data}

\paragraph{Pretraining corpus.} Diagnostic H\&E slides from TCGA (GDC, open access): \her{ov-corpus-selected-slides}
slides across 32 projects, one per case, selected round-robin by project (\her{ov-corpus-selected-brca} TCGA-BRCA).
Of 9,022 indexed slides, 9,015 contributed tiles; 730 were refused (605 RGB JPEG-2000 slides undecodable then,
125 others). Tiling yielded 98,565,315 tiles of $256^2$\,px at 0.5\,\textmu m/px (128\,\textmu m) with tissue fraction
$\geq 0.65$, stored as JPEG (quality 80) in 12,155 shards (2.1\,TB), with a per-slide licence record.

\paragraph{Distillation and JEPA data.} All three training stages draw on one corpus. Feature distillation
reads \emph{teacher caches}: tiles streamed from our corpus in a seeded shuffled order over all shards, each stored with
H-optimus-0's CLS token and its patch tokens average-pooled to a $4\times4$ grid. Our best students' cache (D21) holds
2M tiles, each seen as a single $224^2$ centre view. Later caches vary view (centre, random crop, crop with photometric
jitter) and scale (0.2M--3.2M tiles) for ablations in Supplementary~\ref{app:tried}. Our JEPA stage streams 2.1M corpus
tiles per run (6,000 steps of 352) as centre views, with no augmentation and no teacher outputs. No stage uses HER2 labels, and no Yale-HER2 slide enters any training stage.

\paragraph{Linear-probe set.} Every 100th shard (122 shards), at most 2,000 tiles per project: 64,000 tiles from 454
slides labelled by TCGA project (32 classes; majority 0.073, chance 0.031). Probe tiles come from our corpus, so this probe
measures in-distribution representation quality.

\paragraph{HER2 categories.} Guidelines score HER2 protein by IHC~\citep{wolff2018,wolff2023}: 3+ is
\emph{positive}, 0 and 1+ are \emph{negative}, and an equivocal 2+ goes to ISH, usually fluorescence in situ
hybridisation (FISH), which counts HER2 gene copies; \emph{amplification} makes a 2+ tumour positive. Among negative
tumours, \emph{HER2-low} (1+, or 2+ without amplification) now qualifies for trastuzumab deruxtecan~\citep{modi2022},
and \emph{HER2-zero} (IHC~0) does not. Telling 0 from 1+ is a borderline call: 18 pathologists reading one set of
slides reached 26\% concordance on 0 against 1+, and 58\% on 2+ against 3+~\citep{fernandez2022}.

\paragraph{HER2 cohorts.} \emph{TCGA-BRCA}: \her{ov-cohort-slides} slides (\her{ov-cohort-cases} cases,
\her{ov-cohort-positive} positive), labelled by BCR field \texttt{her2\_status\_by\_ihc} (equivocal and missing
excluded). For part of its positive class that field disagrees with recorded IHC scores: of 156 positives, 82 are
scored 3+, 25 2+, 10 1+ and 39 unscored; 3 of 334 negatives are scored 2+/3+. Slides span at least four native
resolutions. \emph{Yale-HER2} (HER2-TUMOR-ROIS, CC~BY~4.0~\citep{farahmand2022,tcia_her2rois}): 192 slides, 93
positive, one scanner at 0.4986\,\textmu m/px, about one fifth of a TCGA slide's tissue (median 1,196 tiles against
5,832). Every encoder receives identical tiles, at most 400 per slide, stride-sampled across each section (192,600 and
75,108 tiles).

\paragraph{Tissue size in our 490 TCGA slides.} Our 490-slide TCGA-BRCA cohort was assembled one slide per patient,
smallest file first within each class. As a result its negatives carry far less tissue than its positives (median
87 against 134\,mm$^2$ at 1.0\,\textmu m/px), a cue that a classifier can learn and that Yale-HER2 happens to share weakly
(20 against 24\,mm$^2$). It affects every reader trained on these slides
alike, so rankings between encoders stand, but levels on Yale-HER2 are inflated by about 0.04 (\S\ref{sec:res-bias}).
Our reader-level experiments below also train on representative sets for this reason.

\paragraph{More training and test cohorts.} For reader-level experiments we added every labelled TCGA-BRCA diagnostic
slide (931 slides, labels from \texttt{her2\_status\_by\_ihc}, or the guideline rule where it is empty) and
\emph{CPTAC-BRCA}: 395 labelled slides of 120 cases from the Imaging Data Commons~\citep{krug2020cptac} (DICOM whole-slide
series; 382 readable; status by proteogenomic \emph{ERBB2}, clinical where missing). Two further cohorts serve as external
tests only (non-commercial licences): \emph{HEROHE}'s test split~\citep{conde2022herohe}, 150 resection slides (60 positive),
of which 85 are IHC~2+ (32 ISH-amplified), scanned at a stated 0.243\,\textmu m/px; and \emph{BCNB}, 1,058 core-needle
biopsy images (277 positive)~\citep{xu2021bcnb}, with no stated scale (0.465\,\textmu m/px estimated from core width).
HEROHE and BCNB never enter any training; CPTAC-BRCA's tiles enter only R12's re-distillation (\S\ref{sec:res-protocol}), whose
encoders were not adopted.

\paragraph{Borderline and amplification tasks.} Two finer TCGA-BRCA tasks reuse those tiles, with labels from GDC's
clinical fields \texttt{her2\_ihc\_score} and \texttt{her2\_fish\_status}: \emph{HER2-low against HER2-zero} (IHC~1+, or
2+ with negative FISH: 165 slides; IHC~0: 39), and \emph{FISH amplification} among \pp{t4-fish-n} slides with a FISH
result (\pp{t4-fish-pos} amplified). In that FISH-tested subset, FISH agrees with recorded HER2 status for
\pp{t4-fish-agree} of \pp{t4-fish-both} slides.

\subsection{Overlap of our TCGA-BRCA cohort with pretraining}\label{sec:data-overlap}

All \her{ov-overlap-slides} TCGA-BRCA cohort slides are in our pretraining selection, and per-shard licence manifests
show that each contributed tiles ($\approx$3.0M tiles, 3.0\% of the corpus). So for any TCGA-pretrained encoder
(ours, Phikon, Midnight-12k), \TT{} and \YT{} test on slides seen without labels during pretraining, while
\YY{} and \TY{} test on Yale-HER2, external to all our pretraining. No pretraining stage used HER2 labels. We treat
Yale tests as primary evidence.

\section{Methods}\label{sec:methods}

\subsection{HER2 reader}\label{sec:method-overview}

\begin{figure}[tbp]
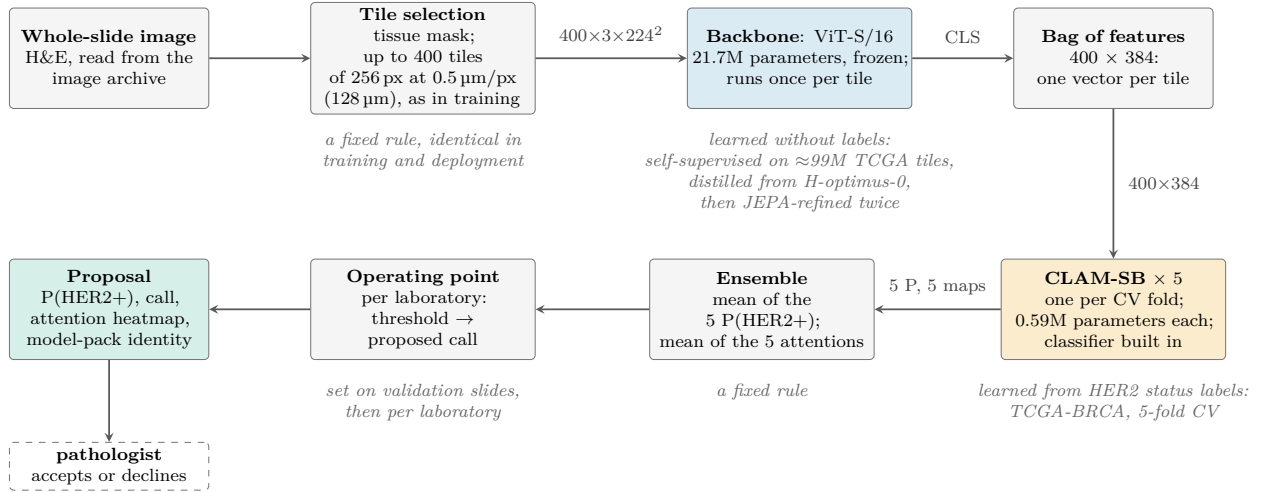

\centering
\resizebox{\linewidth}{!}{\archreaderfigure}
\caption{Our HER2 reader. Blue: tile encoder (self-supervised and distilled, no labels); orange: slide classifiers
(HER2 status labels); grey: fixed rules; green: output for pathologist review. Shapes are per slide.}\label{fig:arch-reader}
\end{figure}

Our reader (Figure~\ref{fig:arch-reader}) turns one H\&E slide into a proposed HER2 status with its evidence, in five
steps. \emph{Tile selection} is identical in training and deployment: up to 400 tissue tiles of $256^2$\,px at
0.5\,\textmu m/px, resized to $224^2$. A frozen \emph{tile encoder}, a ViT-S/16 (12 blocks, width 384, 6 heads; 22M
parameters; Supplementary Table~\ref{tab:backbones}), maps each tile to its final CLS token; ``/16'' denotes 16-pixel patches,
giving 196 patch tokens per tile of $\approx$9\,\textmu m each. A \emph{slide classifier}, CLAM-SB~\citep{lu2021}
implemented from its published equations, pools 400 tile features: a 512-unit projection, gated attention of width
384, attention pooling and a linear classifier ($512\to2$); its attention weights form a heatmap. Five such classifiers,
one per cross-validation fold, make an \emph{ensemble} that averages their probabilities (not logits) and attention
maps. An \emph{operating point} turns a mean probability into a proposed call, which a pathologist accepts or
declines. Only slide classifiers see HER2 labels; our encoder learns from unlabelled tiles
(\S\ref{sec:method-encoder}).

\subsection{Tile encoder: three training stages}\label{sec:method-encoder}

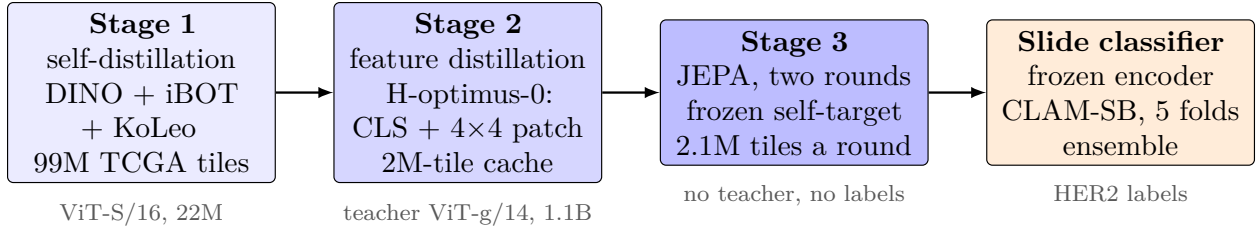
\begin{figure}[tbp]
\centering
\resizebox{\linewidth}{!}{%
\begin{tikzpicture}[font=\small, node distance=4mm and 7mm,
  st/.style={draw, rounded corners=2pt, align=center, minimum height=15mm, text width=30mm, fill=#1},
  ar/.style={-latex, thick}]
\node[st=blue!8] (s1) {\textbf{Stage 1}\\self-distillation\\DINO + iBOT + KoLeo\\99M TCGA tiles};
\node[st=blue!16, right=of s1] (s2) {\textbf{Stage 2}\\feature distillation\\H-optimus-0: CLS + $4{\times}4$ patch\\2M-tile cache};
\node[st=blue!26, right=of s2] (s3) {\textbf{Stage 3}\\JEPA, two rounds\\frozen self-target\\2.1M tiles a round};
\node[st=orange!15, right=of s3] (cl) {\textbf{Slide classifier}\\frozen encoder\\CLAM-SB, 5 folds\\ensemble};
\draw[ar] (s1) -- (s2); \draw[ar] (s2) -- (s3); \draw[ar] (s3) -- (cl);
\node[below=1mm of s1, font=\scriptsize, text=black!65] {ViT-S/16, 22M};
\node[below=1mm of s2, font=\scriptsize, text=black!65] {teacher ViT-g/14, 1.1B};
\node[below=1mm of s3, font=\scriptsize, text=black!65] {no teacher, no labels};
\node[below=1mm of cl, font=\scriptsize, text=black!65] {HER2 labels};
\end{tikzpicture}}
\caption{Training our HER2 reader. Stages 1--3 build our tile encoder from unlabelled public TCGA tiles; only slide
classifiers see labels. Stage 3: a frozen copy of our stage-2 encoder serves as target; a context encoder that sees
20--50\% of patches masked predicts its final-layer tokens at hidden positions.}\label{fig:stages}
\end{figure}

We train our encoder on unlabelled public TCGA tiles in three stages (Figure~\ref{fig:stages}). Stage~1 learns a
general tissue representation by self-supervision; stage~2 transfers what a 1.1B-parameter foundation
model (H-optimus-0) has learned, by feature distillation; stage~3 refines our distilled encoder by predicting its own features of
hidden patches. Our final encoder comes from a second JEPA round on our distilled encoder at step 25,000. We
registered each stage and its decision rule in advance.

\paragraph{Stage 1: self-supervised pretraining.}\label{sec:method-recipe}
\enc{ssl-vits16} is a ViT-S/16~\citep{dosovitskiy2021} with LayerScale~\citep{touvron2021} (21.7M parameters) trained with DINO, iBOT and KoLeo (weight 0.1)
against an EMA teacher, with 1,024 prototypes per head~\citep{caron2021,zhou2022,oquab2024,sablayrolles2019}. It made
one pass over our corpus: 93,338 steps at a global batch of 1,056 (32 per GPU $\times$ 11 accumulation steps $\times$ 3
GPUs). Peak learning rate $1.0155\times10^{-3}$ (square-root scaled), 9,334 warm-up steps, cosine decay to $10^{-6}$;
weight decay 0.04$\to$0.4; EMA momentum 0.994$\to$1.0; teacher temperature 0.04$\to$0.07 over 28,001 steps; 10--50\% of
patches masked with probability 0.5. Each GPU samples tiles from 32 concurrently open shards. Departures from
DINOv2's published recipe were registered before use.

\paragraph{Stain-space augmentation.}\label{sec:method-aug}
DINOv2's photometric augmentation (colour jitter, grayscale, blur, solarisation) makes two views of an H\&E tile nearly
as dissimilar as views of different tiles, and self-distillation then collapses to a uniform prototype assignment
(Supplementary~\ref{app:aug}). We replace it with stain-space jitter: Ruifrok--Johnston deconvolution into haematoxylin,
eosin and residual concentrations~\citep{ruifrok2001}, a per-sample affine perturbation of each concentration~\citep{tellez2019} (standard deviation 0.05), and
reconstruction.

\paragraph{Stage 2: feature distillation.}\label{sec:method-distill}
Starting from a self-supervised encoder, a student learns to reproduce H-optimus-0's representation of each tile.
\emph{Targets}: H-optimus-0's 1,536-d CLS token and its patch tokens average-pooled to a $4\times4$ grid. \emph{Student
heads}: one linear map from student to teacher CLS token, and another from student patch tokens pooled to that grid;
loss is $1-\cos$ on CLS tokens plus, with equal weight, mean $1-\cos$ over 16 grid cells. Both
heads are discarded after training. Teacher outputs are computed once and stored with their tiles in a teacher cache
(\S\ref{sec:data}), so each student step costs one ViT-S forward and backward pass. Recipe (M5.G): batch 64, AdamW~\citep{loshchilov2019}, learning
rate $2\times10^{-4}$, 2,000 warm-up steps, cosine decay~\citep{loshchilov2017} to $10^{-6}$ over 100,000 steps, weight decay 0.04, gradient
clipping 3.0, stain jitter $\sigma=0.05$, on a 2M-tile centre-view cache; snapshots every 25,000 steps. Our ViT-S student
starts from \enc{ssl-vits16} and a ViT-B/16 student from \enc{ssl-vitb16} at step 40,000. Earlier trials matched CLS
tokens only, for 5,000--20,000 steps, with Midnight-12k (trained on TCGA only~\citep{karasikov2025}) or H-optimus-0 as
teacher; students of H-optimus-0 transferred better. In our framework both teachers match their PyTorch references
(cosine 1.000000). H-optimus-0 was trained on a large proprietary collection~\citep{hoptimus2024}, and its students
inherit that provenance.

\paragraph{Stage 3: JEPA refinement.}\label{sec:method-jepa}
A joint-embedding predictive objective~\citep{assran2023,genbio2026} then refines our distilled student. A frozen copy
of our student serves as \emph{target encoder}; a trainable copy, initialised identically, serves as \emph{context
encoder}.
For each tile, 20--50\% of 196 patch positions are hidden (scattered, drawn per sample) and replaced by a learned mask
token at context-encoder input, while our target sees whole tiles. A per-token predictor (LayerNorm,
$D\to2D$, GELU, $2D\to D$) maps context tokens to target tokens, with mean squared error at hidden positions as its
loss. Our target receives no gradient, and we check its hash at every checkpoint; only context weights are kept. Recipe: 6,000 steps of 352 tiles (88 per micro-batch $\times$ 4 accumulation on one GPU; ViT-B: $44\times8$),
AdamW, learning rate $10^{-4}$ with 300 warm-up steps and cosine decay to $10^{-6}$, weight decay 0.04, gradient
clipping 3.0, bf16 trunk with f32 predictor and loss, centre views without augmentation. A \emph{second round} repeats
this stage with round one's result as both target and initialisation. We monitor latent loss relative to a constant
predictor, cosine between predicted and target tokens, and context features' standard deviation (a collapse guard).
Ablations in Supplementary~\ref{app:tried} vary start, learning rate ($3\times10^{-5}$, $2\times10^{-4}$), schedule
(2,000 decayed steps; 12,000; 18,000), photometric augmentation (flip and stain jitter, applied before both encoders),
block masks~\citep{assran2023} and a continuously updated target; each arm was registered before it ran.

\subsection{Slide classifier and operating point}\label{sec:method-head}

CLAM-SB's instance-clustering branch (top and bottom 8 tiles, smooth top-1 SVM, weight 0.3 against 0.7 for slide
loss) is used in training only. Training: Adam $2\times10^{-4}$, weight decay $10^{-5}$, dropout 0.25, at most 200 epochs
with patience 20, 0.59M parameters on ViT-S features. \emph{Cross-validation}: patients, not slides, are dealt into five
class-balanced blocks in an order fixed by a seeded hash of each case identifier (fold seed 1337); the classifier of fold
$k$ trains on three blocks (60\% of patients), keeps the epoch that scores best on block $k{+}1$ and is tested on block $k$,
so each training slide receives one out-of-fold prediction, from a model that never saw it. A separate seed sets
each classifier's initial weights and training order; different seeds on fixed folds measure seed noise
($\approx$0.02 AUC here). Our shipped reader uses five fold models trained on TCGA-BRCA, from a seed fixed in
advance as the lower median of ten by external AUC. We chose its \emph{operating point} without Yale-HER2: maximising
Youden's $J$ on TCGA-BRCA out-of-fold predictions gives a threshold of \pp{rd-thr} on ensemble probability, which we
apply unchanged to new slides. A reader's probabilities shift on a new site, so a laboratory can re-fit that threshold on
its own validated cases or, without labels, normalise scores on its own recent slides. In that case each slide's logit is
standardised over that laboratory's slides, and a threshold set in this scale on training-cohort out-of-fold predictions
is applied (\emph{site normalisation}). For equivocal cases two thresholds can replace one, turning the zone where both
classes overlap into an explicit abstain zone routed to ISH or a second read (Figure~\ref{fig:proto-zones}).

{\booksizes\begin{figure}[!htbp]
\centering
\begin{tikzpicture}
\begin{axis}[width=0.8\textwidth, height=52mm, xmin=-3.5, xmax=5.5, ymin=0, ymax=0.47,
  xlabel={model's HER2 score for a slide}, ylabel={share of slides}, ytick=\empty,
  tick label style={font=\scriptsize}, label style={font=\scriptsize},
  legend style={font=\scriptsize, at={(0.5,-0.28)}, anchor=north, legend columns=2}]
\fill[cdata!10] (axis cs:-3.5,0) rectangle (axis cs:-0.3,0.47);
\fill[cgrey!18] (axis cs:-0.3,0) rectangle (axis cs:1.2,0.47);
\fill[cguard!10] (axis cs:1.2,0) rectangle (axis cs:5.5,0.47);
\addplot[cdata, thick] table[x=x, y=neg] {training/data/auc_dens.dat};
\addplot[cguard, thick] table[x=x, y=p70] {training/data/auc_dens.dat};
\legend{HER2-negative slides, HER2-positive slides}
\node[font=\scriptsize, align=center] at (axis cs:-1.9,0.43) {\textbf{likely negative}\\report, pathologist confirms};
\node[font=\scriptsize, align=center] at (axis cs:0.45,0.43) {\textbf{abstain}\\ISH / second read};
\node[font=\scriptsize, align=center] at (axis cs:3.3,0.43) {\textbf{likely positive}\\report, pathologist confirms};
\end{axis}
\end{tikzpicture}
\caption{\emph{Schematic.} Two thresholds instead of one: the overlap of the two groups becomes an
explicit ``not sure'' zone. The thresholds are set on a validation cohort for a target NPV and PPV, then
carried unchanged to new data (as in the cross-cohort evaluation).}
\label{fig:proto-zones}
\end{figure}
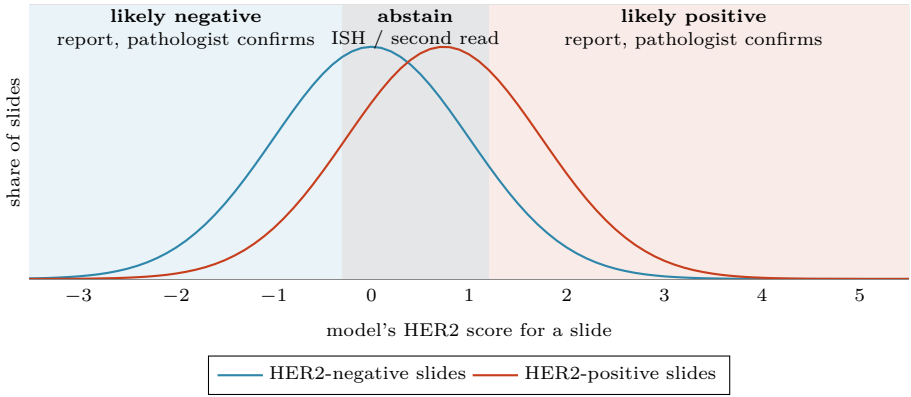}

\subsection{Implementation and deployment}\label{sec:impl}\label{sec:method-deploy}

All code is our own, written on numbat, a deep-learning framework in Zig with a C-ABI SDK whose training stack has been
verified against independent implementations~\citep{tran2026numbat,tran2026crossstack}. Every component passes numeric
gradient, CPU$\leftrightarrow$CUDA equivalence, overfit and bit-exact resume gates before use; forward parity with
PyTorch is cosine $\geq0.9999999$ for our ViT-S and ViT-B trunks, and training throughput exceeds PyTorch's on identical
hardware by 5.3\% for ViT-B/16 (Supplementary~\ref{app:impl}). Training runs data-parallel over three RTX 3090 GPUs.

Our shipped reader travels as one model pack: encoder, five slide classifiers, tile recipe, operating point and an
integrity digest in a single numbat \texttt{.nbq} file, verified before loading. Its encoder matches a PyTorch
re-implementation to $6.3\times10^{-7}$ (maximum absolute difference of tile features); its slide classifiers reproduce
their training predictions to $5\times10^{-8}$; and tiles re-read from fixture slides embed at cosine
$\geq 0.9999999995$ to training bags. A service written on numbat's SDK reads tiles from a department's archive
over DICOMweb, refuses slides coarser than 0.55\,\textmu m/px, and runs as a pod of a Kubernetes (k3s) cluster on a
GPU-free node (12-core x86-64 CPU, 32\,GB RAM) shared with other services of a pathology platform. On three fixture
slides read from that archive, our deployed reader returns ensemble probabilities within $6\times10^{-8}$ of training
predictions, in 5.2--5.6\,s per slide. It is for research use only.

\paragraph{In the pathology platform.} This platform profiles breast tumour markers for pathologists. It already scores
nuclear IHC markers (ER, PR and Ki-67), and our HER2 reader adds an H\&E-based HER2 proposal beside them.
Figure~\ref{fig:f-cv-reader} follows one slide through our deployed reader. Up to 400 tiles are encoded once; each of
five fold classifiers reads that bag and gives a probability; their mean is compared with a local operating point; and
a pre-read shows probability, proposed call and operating point to a pathologist, who decides and signs. Encoding takes
almost all of the cost (about 1.8 trillion multiply-adds per slide), and five classifiers add under 0.1\%. Our reader
proposes a status from H\&E; scoring the HER2 IHC slide stays with the pathologist.

{\booksizes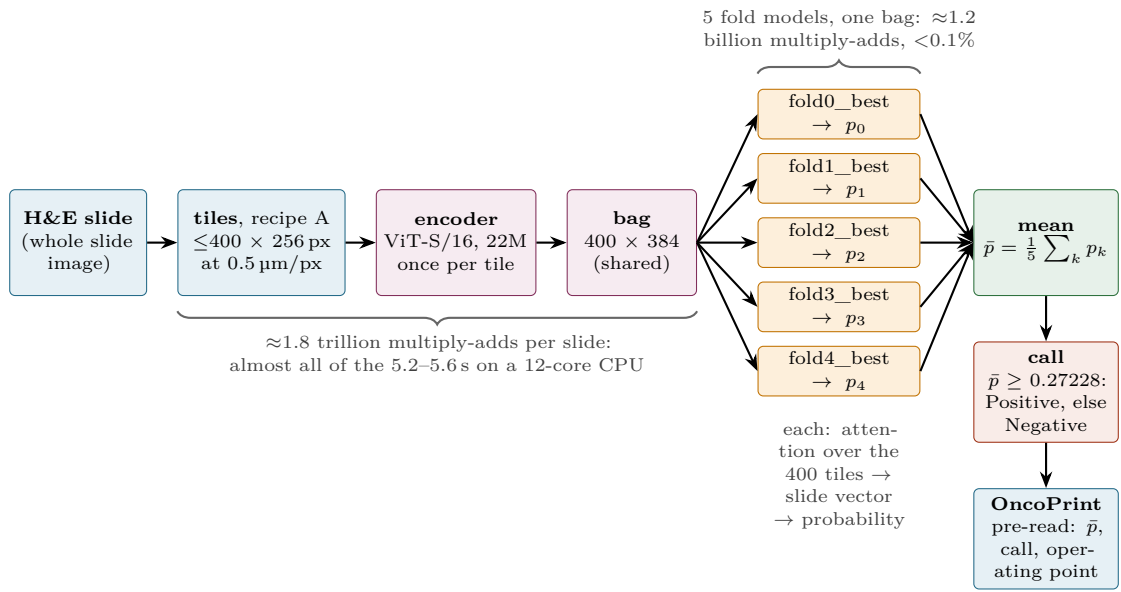
\begin{figure}[!htbp]
\centering
\begin{tikzpicture}[font=\scriptsize, box/.append style={font=\scriptsize}]
\node[dbox, text width=16mm, minimum height=14mm] (slide) {\textbf{H\&E slide}\\(whole slide\\image)};
\node[dbox, text width=20mm, minimum height=14mm, right=4mm of slide] (tiles) {\textbf{tiles}, recipe A\\$\le$400 $\times$ 256\,px\\at 0.5\,\textmu m/px};
\node[mbox, text width=19mm, minimum height=14mm, right=4mm of tiles] (enc) {\textbf{encoder}\\ViT-S/16, 22M\\once per tile};
\node[mbox, text width=15mm, minimum height=14mm, right=4mm of enc] (bag) {\textbf{bag}\\400 $\times$ 384\\(shared)};
\foreach \k in {0,...,4} {
  \node[gbox, text width=20mm, minimum height=6mm, inner sep=2pt, right=8mm of bag, yshift=(2-\k)*8.5mm] (h\k) {fold\k\_best $\to p_{\k}$};
  \draw[arr] (bag.east) -- (h\k.west);
}
\node[okbox, text width=17mm, minimum height=14mm, right=7mm of h2] (avg) {\textbf{mean}\\$\bar p = \frac{1}{5}\sum_k p_k$};
\foreach \k in {0,...,4} \draw[arr] (h\k.east) -- (avg.west);
\node[rbox, text width=17mm, minimum height=12mm, below=6mm of avg] (call) {\textbf{call}\\$\bar p \ge 0.27228$:\\Positive, else Negative};
\node[dbox, text width=17mm, minimum height=12mm, below=6mm of call] (op) {\textbf{OncoPrint}\\pre-read: $\bar p$,\\call, operating point};
\draw[arr] (slide) -- (tiles); \draw[arr] (tiles) -- (enc); \draw[arr] (enc) -- (bag); \draw[arr] (avg) -- (call); \draw[arr] (call) -- (op);
\draw[decorate, decoration={brace, amplitude=5pt, mirror}, thick, black!60] ([yshift=-2mm]tiles.south west) -- ([yshift=-2mm]bag.south east)
  node[midway, below=5pt, lbl, text width=58mm, align=center]{$\approx$1.8 trillion multiply-adds per slide:\\almost all of the 5.2--5.6\,s on a 12-core CPU};
\draw[decorate, decoration={brace, amplitude=5pt}, thick, black!60] ([yshift=2mm]h0.north west) -- ([yshift=2mm]h0.north east)
  node[midway, above=5pt, lbl, text width=40mm, align=center]{5 fold models, one bag: $\approx$1.2 billion multiply-adds, $<$0.1\%};
\node[lbl, text width=24mm, align=center, below=2mm of h4] {each: attention over the\\400 tiles $\to$ slide vector\\$\to$ probability};
\end{tikzpicture}
\caption{\emph{Exact} (the shipped pack's manifest and reader) with \emph{measured} times. The five fold models of
cross-validation are kept and read together: the encoder runs once, each fold model reads the same bag, and their five
probabilities are averaged before the threshold.}\label{fig:f-cv-reader}
\end{figure}}

\subsection{Evaluation}\label{sec:method-eval}

\begin{figure}[tbp]
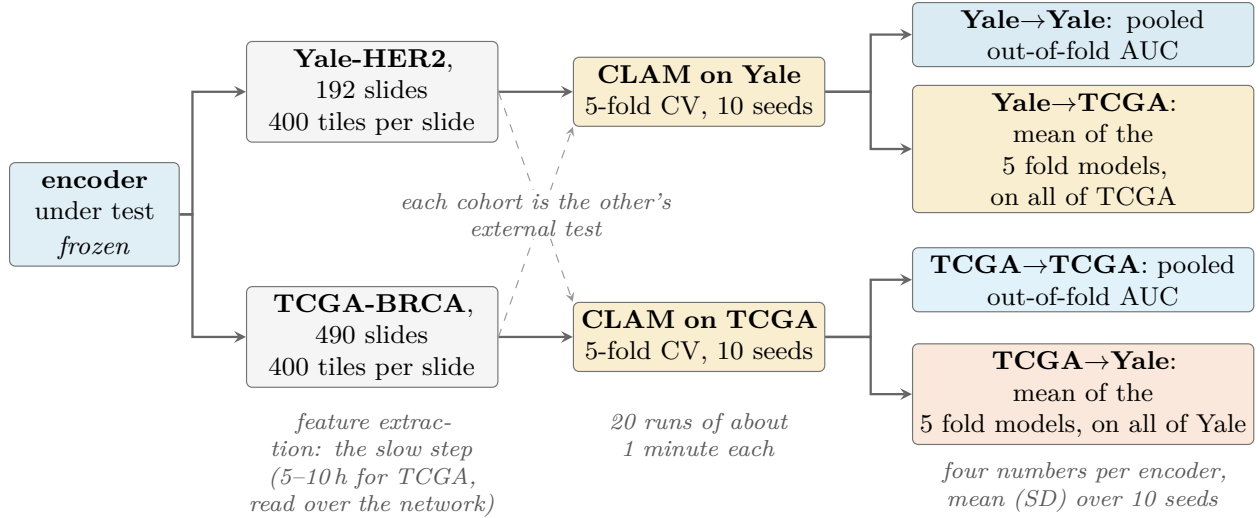

\centering
\resizebox{\linewidth}{!}{\herprotocolfigure}
\caption{HER2 evaluation of one frozen encoder: feature extraction once per cohort, then slide-classifier runs for 2
training cohorts $\times$ 10 seeds, each 5-fold cross-validated, yield four tests.}\label{fig:protocol}
\end{figure}

\paragraph{HER2 status: readers that differ only in their encoder.} For each encoder (Figure~\ref{fig:protocol}),
400-tile bags of both cohorts are extracted once; slide classifiers are trained on each cohort under 5-fold
case-stratified cross-validation for ten seeds, and each seed's five fold models are applied to a second cohort, which
never enters any split. That gives four tests: \YY{} and \TT{} (within a cohort; AUC of pooled out-of-fold predictions)
and \TY{} and \YT{} (across cohorts; AUC of a five-model ensemble on that second cohort). Each test is a mean over ten
seeds. Seed $s$ uses identical folds for every encoder, so differences are paired: we report mean per-seed
difference, seeds in its favour ($k/10$) and an exact one-sided Wilcoxon signed-rank $p$. All runs share one
software build (numbat SDK 0.23.0), since CLAM is deterministic per build. \TY{} is our primary test: Yale-HER2 is
external to all our training (\S\ref{sec:data-overlap}).

\paragraph{HER2 status: a local PathBench protocol.} PathBench ranks encoders on private cohorts held out of all
pretraining~\citep{pathbench2025}, so its rankings cannot be rerun or extended on public data. We rebuilt its
classification protocol on our two public cohorts: an ABMIL classifier~\citep{ilse2018} trained from scratch for each
task, encoder and seed; one case-stratified 7:1:2 train/validation/test split per seed, seeds 1--10, shared by every
encoder; test AUC with a 1,000-sample bootstrap; and exact one-sided paired Wilcoxon tests against each cohort's best
encoder. Before using it we required that it reproduce PathBench's published order of reference encoders on breast HER2
(H-optimus-0 $>$ Phikon $>$ Hibou $>$ ResNet-50). Fresh classifiers and held-out test splits make it independent of our
first protocol's cross-validation and cross-cohort transfer, so each protocol checks the other (Supplementary~\ref{app:pb}).

\paragraph{Readers on new hospitals.} Here we judge a reader as a whole (tiles, encoder, classifier, operating point) on
hospitals it never trained on: Yale-HER2, HEROHE (with its IHC~2+ subset scored on its own: ISH-amplified against not)
and BCNB, trained on TCGA-BRCA or on Yale-HER2, ten seeds, mean per-seed AUC, and sensitivity and specificity at the
training cohort's out-of-fold Youden point, raw and site-normalised. We registered each experiment with its decision
rule before running it, and adopted a change only on a paired gain larger than seed noise. Four tile recipes were compared:
\emph{A}, our default (at most 400 tiles of 128\,\textmu m at 0.5\,\textmu m/px); \emph{C}, at most 2,000 such tiles;
\emph{B} and \emph{D}, at most 2,000 or 400 tiles of 224\,\textmu m at 1.0\,\textmu m/px. Training sets ranged from our
490 TCGA slides to all 931 and to TCGA plus CPTAC-BRCA, with diagnostic subsets of the 490 (Supplementary~\ref{app:reader}).
\emph{The published reader's protocol} was replicated with our encoder: every tissue tile at 1.0\,\textmu m/px (recipe
\emph{E}; median 1,835 tiles per slide), attention MIL with non-gated attention (CLAM without its instance branch), 5 folds
$\times$ 8 seeds averaged into one 40-model reader, trained on TCGA (917 slides) plus CPTAC-BRCA (382), our pinned
optimiser where the publication is silent~\citep{athena2025}. \emph{R12} distilled H-optimus-0 again into our encoder at
reading scale: 1.1M new tiles at 1.0\,\textmu m/px from TCGA and CPTAC-BRCA, a 1.4M-record teacher cache (1.0M at
1.0\,\textmu m/px, 0.4M at 0.5), 25,000 distillation steps from our final encoder with the stage-2 recipe, then one JEPA
round; each candidate was read under the published protocol and adopted only if its Yale AUC reached 0.829 (ours under
that protocol $+0.02$) and beat ours on at least 6 of 8 paired seeds.

\paragraph{Borderline and amplification tasks.} \emph{HER2-low against HER2-zero} and \emph{FISH amplification} are read
in two ways on TCGA-BRCA. One is a \emph{dedicated classifier}: CLAM trained on each task's own labels with identical
settings (5-fold cross-validation, ten seeds, pooled out-of-fold AUC). The other is a \emph{HER2-status reader} used as
it is, with its HER2 probability as score: readers trained on Yale-HER2 and applied to TCGA-BRCA (never seen), and
readers trained on TCGA-BRCA scored out of fold. Neither way of using a status reader ever sees a FISH or IHC-score
label.

\paragraph{CPU cost.} Tile-encoder time per 400-tile slide on a GPU-free 12-core server of a department cluster
(32\,GB RAM), for three architectures (ViT-S/16, ViT-B/16, ViT-g/14), with one library, f32, estimated as a slope
between two tile counts so that model loading cancels (Supplementary~\ref{app:cpu}).

\section{Results}\label{sec:results}

\subsection{HER2 status: readers that differ only in their encoder}\label{sec:res-rank}

\begin{table}[tbp]
\centering\footnotesize
\caption{HER2 status from readers that differ only in their tile encoder, ranked by external validation (\TY).}\label{tab:rank}
\begin{threeparttable}
\begin{tabularx}{\linewidth}{@{}>{\raggedright\arraybackslash}X l r r cccc@{}}
\toprule
tile encoder & architecture & params & CPU s & Yale$\to$Yale & TCGA$\to$TCGA & TCGA$\to$Yale & Yale$\to$TCGA \\
\midrule
\textbf{Ours (final)} & ViT-S/16 & 22M & 8.4 & \textbf{0.844$^{3}$} & \textbf{0.690$^{1}$} & \textbf{0.812$^{1}$} & \textbf{0.650$^{3}$} \\
Phikon & ViT-B/16 & 86M & 23.0 & 0.872$^{2}$ & 0.679$^{2}$ & 0.811$^{2}$ & 0.662$^{1}$ \\
H-optimus-0 & ViT-g/14 & 1.1B & 312 & 0.897$^{1}$ & 0.676$^{3}$ & 0.805$^{3}$ & 0.653$^{2}$ \\
Hibou-B & ViT-B/14 & 86M & -- & 0.786$^{5}$ & 0.626$^{4}$ & 0.749$^{4}$ & 0.637$^{5}$ \\
Midnight-12k & ViT-g/14 & 1.1B & 312$^\dagger$ & 0.834$^{4}$ & 0.623$^{5}$ & 0.727$^{5}$ & 0.638$^{4}$ \\
ResNet-50$^\ast$ & CNN & 26M & -- & 0.700$^{6}$ & 0.545$^{6}$ & 0.638$^{6}$ & 0.568$^{7}$ \\
DINOv2-S$^\ast$ & ViT-S/14 & 22M & -- & 0.676$^{7}$ & 0.495$^{7}$ & 0.589$^{7}$ & 0.573$^{6}$ \\
\bottomrule
\end{tabularx}

\begin{tablenotes}[flushleft]\footnotesize
\item[] AUC, mean over 10 paired seeds. Superscript: rank in its column. CPU s: seconds to encode one 400-tile slide on a GPU-free 12-core server; --, not timed. Standard deviations in Supplementary Table~\ref{tab:v23}.
\item[$^\dagger$] Timed on an identical architecture, H-optimus-0's ViT-g/14.
\item[$^\ast$] Natural-image controls (ImageNet ResNet-50; DINOv2 ViT-S/14).
\end{tablenotes}
\end{threeparttable}
\end{table}

\begin{figure}[tbp]
\centering
\begin{minipage}[t]{0.56\linewidth}\centering
\begin{tikzpicture}
\begin{axis}[xbar, bar shift=0pt, width=0.55\linewidth, height=58mm, bar width=6pt, xmin=0.55, xmax=0.86,
  ytick=data, yticklabels from table={data/rank_ty.dat}{label}, y tick label style={font=\scriptsize},
  x tick label style={font=\scriptsize, /pgf/number format/fixed, /pgf/number format/precision=2},
  xlabel={AUC, TCGA$\to$Yale}, label style={font=\small}, enlarge y limits=0.08,
  title={\small (a) ranking}]
\addplot[fill=black!25, draw=black!50, error bars/.cd, x dir=both, x explicit]
  table[x=ty, y=y, x error=sd] {data/rank_ty.dat};
\addplot[fill=red!65!black, draw=black, error bars/.cd, x dir=both, x explicit]
  table[x=ty, y=y, x error=sd] {data/rank_ty_own.dat};
\end{axis}
\end{tikzpicture}
\end{minipage}\hfill
\begin{minipage}[t]{0.40\linewidth}\centering
\begin{tikzpicture}
\begin{axis}[xmode=log, width=0.92\linewidth, height=58mm, xmin=12, xmax=2500, ymin=0.56, ymax=0.84,
  xlabel={encoder parameters (millions, log)}, ylabel={AUC}, label style={font=\small},
  tick label style={font=\scriptsize}, y tick label style={/pgf/number format/fixed, /pgf/number format/precision=2},
  grid=major, grid style={black!10}, title={\small (b) accuracy against size}]
\addplot[only marks, mark=*, mark size=2pt, black!55] table[x=params, y=ty] {data/rank_ty.dat};
\addplot[only marks, mark=star, mark size=4.5pt, red!65!black, thick] table[x=params, y=ty] {data/rank_ty_own.dat};
\node[font=\scriptsize, text=red!65!black, anchor=north west] at (axis cs:23,0.808) {ours};
\node[font=\tiny, anchor=north west] at (axis cs:86,0.808) {Phikon};
\node[font=\tiny, anchor=south east] at (axis cs:1100,0.809) {H-optimus-0};
\node[font=\tiny, anchor=north west] at (axis cs:86,0.749) {Hibou-B};
\node[font=\tiny, anchor=north east] at (axis cs:1100,0.727) {Midnight-12k};
\node[font=\tiny, anchor=west] at (axis cs:27,0.638) {ResNet-50};
\node[font=\tiny, anchor=west] at (axis cs:23,0.589) {DINOv2-S};
\end{axis}
\end{tikzpicture}
\end{minipage}
\caption{HER2 status, readers trained on TCGA-BRCA and tested on Yale-HER2 (external validation; 10 paired seeds, bars:
SD). \textbf{(a)} Ranking by tile encoder. \textbf{(b)} AUC against encoder size; star: our reader.}\label{fig:rank}
\end{figure}
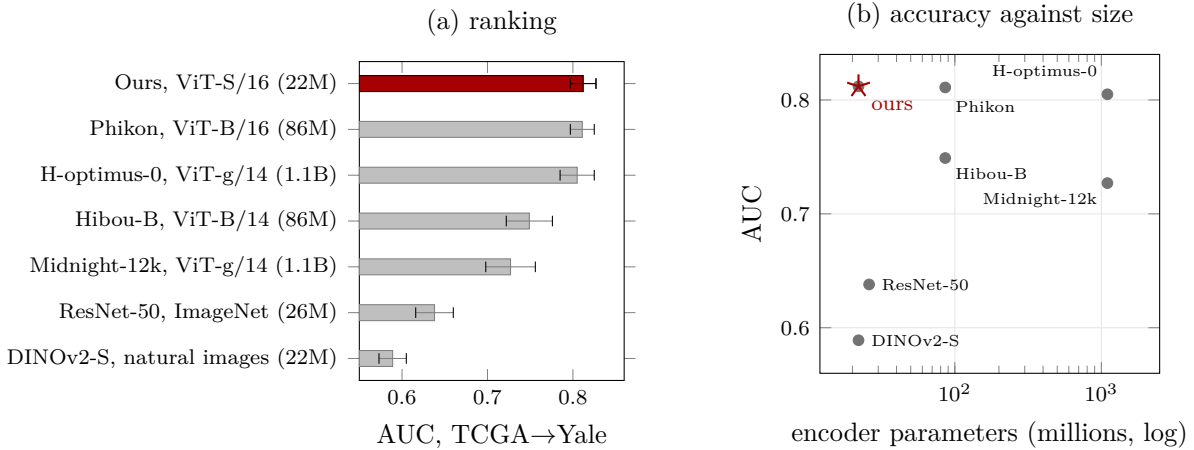

With every other part held fixed, our encoder gives the best reader of \pp{rank-n} on both tests trained on TCGA-BRCA
and the third best on both tests trained on Yale-HER2 (Table~\ref{tab:rank}, Figure~\ref{fig:rank}). On external
validation (\TY) our reader reaches \mm{jj25s3-ty}, level with a reader on Phikon (\mm{phikon-ty}; paired difference
\mm{pd-jj25s3-phikon-ty}, \mm{pd-jj25s3-phikon-ty-k} seeds) and just ahead of one on H-optimus-0 (\mm{hopt-ty}; ahead on
\mm{pd-jj25s3-hopt-ty-k} seeds). On TCGA-BRCA (\TT) it leads Phikon's by \mm{pd-jj25s3-phikon-tt}
(\mm{pd-jj25s3-phikon-tt-k}, $p=\mm{pd-jj25s3-phikon-tt-p}$). On both Yale-trained tests it ranks third, after readers on
H-optimus-0 and Phikon. Across four tests our reader takes first place twice, more often than any other reader here,
with an encoder a quarter of Phikon's size and a fiftieth of H-optimus-0's. Readers on Hibou-B (86M) and
Midnight-12k (1.1B) rank below ours on every test, and both natural-image controls rank last. Two more runs of our full
encoder recipe, from independent seeds, reach \mm{jj25s4-ty} and \mm{jj25s5-ty} on \TY{} (Supplementary~\ref{app:v23}).
All readers in this comparison train on one set of 490 TCGA-BRCA slides, whose size bias (\S\ref{sec:data}) inflates each
Yale level by about 0.04 (\S\ref{sec:res-bias}). Since every encoder shares that bias, comparisons between encoders stay
paired, but absolute levels sit above what a representative training set gives.

\subsection{A second protocol: local PathBench}\label{sec:res-pb}

\begin{table}[tbp]
\centering\footnotesize
\caption{HER2 status under our local PathBench protocol (Supplementary~\ref{app:pb}).}\label{tab:pb-rank}
\begin{threeparttable}
\begin{tabularx}{\linewidth}{@{}>{\raggedright\arraybackslash}X l r cc cc@{}}
\toprule
 & & & \multicolumn{2}{c}{TCGA-BRCA} & \multicolumn{2}{c}{Yale-HER2} \\
\cmidrule(lr){4-5}\cmidrule(lr){6-7}
tile encoder & architecture & params & AUC & $p$ vs best & AUC & $p$ vs best \\
\midrule
\textbf{Ours (final)} & ViT-S/16 & 22M & \textbf{0.689$^{1}$} & -- & \textbf{0.897$^{3}$} & 0.16 \\
H-optimus-0 & ViT-g/14 & 1.1B & 0.660$^{2}$ & 0.08 & 0.913$^{1}$ & -- \\
Phikon & ViT-B/16 & 86M & 0.656$^{3}$ & 0.18 & 0.901$^{2}$ & 0.18 \\
Midnight-12k & ViT-g/14 & 1.1B & 0.630$^{4}$ & $<$0.01 & 0.878$^{4}$ & 0.02 \\
Hibou-B & ViT-B/14 & 86M & 0.605$^{5}$ & $<$0.01 & 0.853$^{5}$ & $<$0.01 \\
DINOv2-S$^\ast$ & ViT-S/14 & 22M & 0.541$^{7}$ & $<$0.01 & 0.794$^{6}$ & $<$0.01 \\
ResNet-50$^\ast$ & CNN & 26M & 0.567$^{6}$ & $<$0.01 & 0.730$^{7}$ & $<$0.01 \\
\bottomrule
\end{tabularx}

\begin{tablenotes}[flushleft]\footnotesize
\item[] A fresh ABMIL classifier per task, encoder and seed; held-out test splits. Test AUC, mean over 10 paired seeds. Superscript: rank in its column. $p$: exact one-sided Wilcoxon test that the column's best encoder is ahead.
\item[$^\ast$] Natural-image controls.
\end{tablenotes}
\end{threeparttable}
\end{table}

Our local PathBench protocol (\S\ref{sec:method-eval}) passes its check: on both public cohorts it orders PathBench's
reference encoders exactly as PathBench did on its private cohort (H-optimus-0 $>$ Phikon $>$ Hibou $>$ ResNet-50), so
encoders can now be ranked for HER2 on open data under PathBench's rules. Under it our encoder ranks first of \pp{pb-n}
on TCGA-BRCA (AUC \pp{pb-tcga-ours}), ahead of H-optimus-0 in 7 of 10 paired seeds, and third on Yale-HER2
(\pp{pb-yale-ours}; Table~\ref{tab:pb-rank}). On both cohorts it is statistically level with Phikon and H-optimus-0
($p\geq0.08$) and ahead of Midnight-12k (1.1B), Hibou-B and both controls. Ranked among every readout of every
foundation model we scored, it stays first of 14 on TCGA-BRCA, and no model is ahead of it at $p<0.05$ on either cohort
(Supplementary Figure~\ref{fig:pb-rank}). Two independent protocols thus agree: our 22M-parameter encoder sits in one
group with models 4 and 50 times its size.

\subsection{What each training stage contributes}\label{sec:res-stages}

\begin{table}[tbp]
\centering\footnotesize
\caption{Readers on our ViT-S/16 after each encoder training stage.}\label{tab:stages}
\begin{threeparttable}
\begin{tabularx}{\linewidth}{@{}>{\raggedright\arraybackslash}X cccc@{}}
\toprule
 & \YY & \TT & \TY & \YT \\
\midrule
Stage 1: self-supervised pretraining & \mm{sfinal-yy} & \mm{sfinal-tt} & \mm{sfinal-ty} & \mm{sfinal-yt} \\
Stage 2: + distillation of H-optimus-0 & \mm{m5s25-yy} & \mm{m5s25-tt} & \mm{m5s25-ty} & \mm{m5s25-yt} \\
Stage 3: + JEPA, round 1 & \pp{r1-yy} & \pp{r1-tt} & \pp{r1-ty} & \pp{r1-yt} \\
Stage 3: + JEPA, round 2 (final) & \textbf{\mm{jj25s3-yy}} & \textbf{\mm{jj25s3-tt}} & \textbf{\mm{jj25s3-ty}} & \textbf{\mm{jj25s3-yt}} \\
\midrule
Phikon (reference) & \mm{phikon-yy} & \mm{phikon-tt} & \mm{phikon-ty} & \mm{phikon-yt} \\
\bottomrule
\end{tabularx}
\begin{tablenotes}[flushleft]\footnotesize
\item[] AUC, mean over 10 paired seeds. Stage 2: distilled encoder at step 25,000. JEPA round 1: mean of three seeds.
\end{tablenotes}
\end{threeparttable}
\end{table}

\begin{figure}[tbp]
\centering
\begin{tikzpicture}
\begin{groupplot}[group style={group size=4 by 1, horizontal sep=9mm}, width=0.29\linewidth, height=46mm,
  xtick={0,1,2,3}, xticklabels={SSL, distil, JEPA 1, JEPA 2}, x tick label style={font=\scriptsize, rotate=40,
  anchor=north east}, y tick label style={font=\scriptsize, /pgf/number format/fixed, /pgf/number format/precision=2},
  title style={font=\small}, xmin=-0.3, xmax=3.3, every axis plot/.append style={thick}]
\input{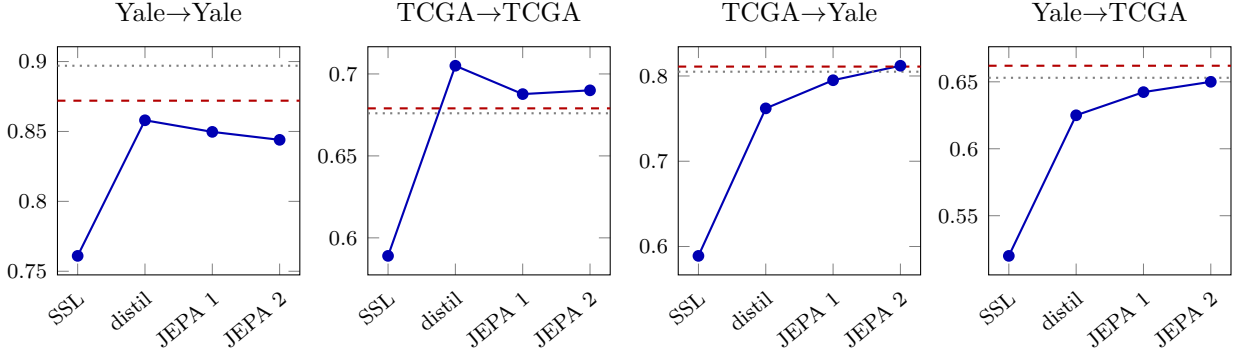}
\end{groupplot}
\end{tikzpicture}
\caption{Readers on our ViT-S/16 after each encoder training stage (10 paired seeds). Dashed: Phikon (ViT-B/16, 86M);
dotted: H-optimus-0 (ViT-g/14, 1.1B), our distillation teacher.}\label{fig:ladder}
\end{figure}

Self-supervised pretraining alone gives a weak cross-cohort reader (\TY{} \mm{sfinal-ty}). Distilling H-optimus-0 lifts
every test, taking \TY{} to \mm{m5s25-ty} (Table~\ref{tab:stages}, Figure~\ref{fig:ladder}). A first JEPA round adds
\mm{pd-j25s4-m5s25-ty} to \mm{pd-j25s3-m5s25-ty} on \TY{} across three seeds (9--10 of 10 paired seeds each) and about
+0.017 on \YT; a second round, with its target refreshed, adds \mm{pd-jj25s3-j25s3-ty} more on \TY{} and
\mm{pd-jj25s3-j25s3-yt} on \YT. JEPA gains on cross-cohort tests only; within a cohort, two rounds cost
\mm{pd-jj25s3-m5s25-yy} on \YY{} and \mm{pd-jj25s3-m5s25-tt} on \TT. Runs from one start agree closely: three
first-round seeds lie within \pp{r1-ty-spread} of one another on every test, and three full two-round runs within 0.002 on
\TY. Supplementary Table~\ref{tab:tried} lists every other change we tested at each stage.

\subsection{Our shipped reader at its operating point}\label{sec:res-reader}

\begin{table}[tbp]
\centering\footnotesize
\caption{Our shipped reader at its operating point and at 0.5.}\label{tab:reader}
\begin{threeparttable}
\begin{tabularx}{\linewidth}{@{}>{\raggedright\arraybackslash}X l cccc@{}}
\toprule
cohort & threshold & sensitivity & specificity & PPV & NPV \\
\midrule
TCGA-BRCA, out of fold (\pp{rd-n-tcga} slides) & \pp{rd-thr} & \pp{rd-tcga-sens} & \pp{rd-tcga-spec} & \pp{rd-tcga-ppv} & \pp{rd-tcga-npv} \\
Yale-HER2, external (\pp{rd-n-yale} slides) & \pp{rd-thr} & \textbf{\pp{rd-yale-sens}} [\pp{rd-yale-sens-ci}] & \textbf{\pp{rd-yale-spec}} [\pp{rd-yale-spec-ci}] & \pp{rd-yale-ppv} & \pp{rd-yale-npv} \\
Yale-HER2, external & 0.5 & \pp{rd-yale05-sens} [\pp{rd-yale05-sens-ci}] & \pp{rd-yale05-spec} [\pp{rd-yale05-spec-ci}] & \pp{rd-yale05-ppv} & \pp{rd-yale05-npv} \\
\bottomrule
\end{tabularx}
\begin{tablenotes}[flushleft]\footnotesize
\item[] Five TCGA-BRCA-trained slide classifiers on our final encoder. Threshold \pp{rd-thr} on their ensemble probability, fixed by Youden's $J$ on TCGA-BRCA out-of-fold predictions and applied unchanged to Yale-HER2. Brackets: Wilson 95\% intervals.
\end{tablenotes}
\end{threeparttable}
\end{table}

A reader proposes a call, so its threshold matters as much as its AUC. Our shipped reader uses one seed's five
TCGA-trained slide classifiers, chosen by a rule fixed in advance (\S\ref{sec:method-head}); on Yale-HER2 their ensemble
reaches AUC \pp{rd-auc} (DeLong 95\% interval \pp{rd-auc-ci}~\citep{delong1988}). On a new site its probabilities run
low: at 0.5 it finds only \pp{rd-yale05-sens} of HER2-positive slides (Table~\ref{tab:reader}). A threshold fixed on
TCGA alone (\pp{rd-thr}) transfers well, giving sensitivity \pp{rd-yale-sens} and specificity \pp{rd-yale-spec} on
Yale-HER2, higher on both counts than out of fold on TCGA-BRCA, whose labels are noisier (\S\ref{sec:data}). A
department confirms its threshold on its own validated cases before use (\S\ref{sec:res-bias}).

\subsection{Our reader on three external hospitals, and on equivocal IHC~2+}\label{sec:res-hospitals}

\paragraph{Tile scale.} On our 490 TCGA slides, 400 tiles of 224\,\textmu m at 1.0\,\textmu m/px (recipe D) raised Yale
AUC from 0.81 to 0.85 over our default recipe A (9 of 10 paired seeds), while 2,000 tiles at either scale did not
(Supplementary~\ref{app:reader}). The gain did not travel: on BCNB, D trailed A by 0.02 (TCGA-trained) and 0.03
(Yale-trained), on HEROHE by 0.01 and 0.03, and on HEROHE's IHC~2+ cases by 0.02 and 0.03. We kept recipe A.

\paragraph{Three hospitals.} Our deployed recipe reads BCNB's biopsies at 0.63 when trained on TCGA-BRCA and 0.70 when
trained on Yale-HER2, and HEROHE at 0.75 and 0.84 (Table~\ref{tab:proto-measured}). Which hospital supplies training
slides matters more than any recipe we tried: Yale-trained readers lead TCGA-trained ones by 0.06--0.10 on both. On an
identical pipeline, a reader on H-optimus-0 leads ours on BCNB and HEROHE by about 0.02 when both train on TCGA-BRCA, is
level on Yale, and trails ours on HEROHE when both train on Yale-HER2.

\paragraph{The equivocal IHC~2+ cases.} Among HEROHE's 85 IHC~2+ tumours, 32 ISH-amplified, our deployed reader ranks
amplified above non-amplified tumours at AUC 0.76 when trained on TCGA-BRCA and 0.79 when trained on Yale-HER2
(Figure~\ref{fig:her2-hard}), against 0.78 and 0.80 for a reader on H-optimus-0 and 0.82 for the best published figure on
these cases (H-optimus-1 with CLAM, under its own protocol~\citep{lv2026}). Every other training set and protocol we tried
ranks them lower (0.60--0.74). On these cases our reader orders ISH requests by likely amplification and flags
discordant ones, while ISH settles them.

{\booksizes\begin{table}[!htbp]
\centering\scriptsize\setlength{\tabcolsep}{3pt}
\caption{\emph{Measured} HER2 AUC on our cohorts, with the hard cases first.}\label{tab:proto-measured}
\begin{threeparttable}
\begin{tabularx}{\textwidth}{@{}>{\raggedright\arraybackslash}Xcccccc>{\raggedright\arraybackslash}p{11mm}@{}}
\toprule
 & equivocal & \multicolumn{2}{c}{IHC 2+, HEROHE} & \multicolumn{3}{c}{status} & licence \\
\cmidrule(lr){3-4}\cmidrule(lr){5-7}
reader (encoder) & TCGA & TCGA- & Yale- & Yale & TCGA & TCGA \\
 & 2+$\to$ISH & trained & trained & $\to$Yale & $\to$TCGA & $\to$Yale & \\
\midrule
\multicolumn{8}{@{}l}{\emph{released encoders, same pipeline (CLAM-SB, 400 tiles at 0.5\,\textmu m/px)}} \\
Virchow & \textbf{0.758} & --- & --- & --- & --- & --- & Apache-2.0 \\
Hibou-B & 0.747 & --- & --- & \her{hibou-yy} & \her{hibou-tt} & \her{hibou-ty} & Apache-2.0 \\
H-optimus-0 & 0.738 & \textbf{0.777} & \textbf{0.795} & \textbf{\her{hoptaug-yy}} & \her{hoptaug-tt} & \her{hoptaug-ty} & Apache-2.0 \\
Phikon v1 & 0.709 & --- & --- & \her{phikon-yy} & \her{phikon-tt} & \textbf{\her{phikon-ty}} & non-comm. \\
ResNet-50 (natural images) & 0.676 & --- & --- & \her{rn50-yy} & \her{rn50-tt} & \her{rn50-ty} & --- \\
\midrule
\multicolumn{8}{@{}l}{\emph{our readers (ViT-S/16, 22M)}} \\
\textbf{deployed}: CLAM-SB, 400 tiles, 0.5\,\textmu m/px, TCGA 490 & --- & \textbf{0.761} & \textbf{0.789} & \mm{jj25s3-yy} & \mm{jj25s3-tt} & \mm{jj25s3-ty} & owned \\
staged D pack: 400 tiles at 1.0\,\textmu m/px & --- & 0.743 & 0.759 & --- & --- & 0.851 & owned \\
R10: TCGA 927 (representative), 0.5\,\textmu m/px & --- & 0.713 & --- & --- & --- & 0.801 & owned \\
R11-O: published protocol & --- & 0.596 & --- & --- & --- & 0.809$^b$ & owned \\
R12 A-25k: re-distilled at 1.0\,\textmu m/px & --- & 0.658 & --- & --- & --- & 0.813$^b$ & owned \\
R12 B-6k: A-25k + JEPA & --- & 0.659 & --- & --- & --- & 0.812$^b$ & owned \\
\midrule
\multicolumn{8}{@{}l}{\emph{published (other protocols and training sets)}} \\
H-optimus-1 + CLAM (Lv et al.) & --- & \multicolumn{2}{c}{0.817$^c$} & --- & --- & 0.900$^c$ & --- \\
H-optimus-0 + ABMIL (Bosch et al.) & --- & --- & --- & --- & --- & 0.907$^b$ & Apache-2.0 \\
\bottomrule
\end{tabularx}
\begin{tablenotes}[flushleft]\scriptsize
\item[] \emph{Equivocal, TCGA}: pooled out-of-fold AUC on the 169 TCGA-BRCA IHC~2+ slides for ISH amplification (clean-room CLAM, 5 seeds, August).
\item[] \emph{IHC 2+, HEROHE}: the external HEROHE test split's 85 IHC~2+ slides (32 amplified); readers trained on TCGA-BRCA or on Yale-HER2; per-seed mean (R9, R11, R12).
\item[] \emph{Status}: in-cohort pooled out-of-fold AUC and the TCGA$\to$Yale five-model ensemble (released encoders: declared metric, 3 seeds; ours: 10 seeds, v23 protocol). ---, not measured.
\item[$^b$] The published reader's protocol (TCGA + CPTAC, all tiles, 40 models).
\item[$^c$] Their protocol and training set (HEROHE-IHC2).
\end{tablenotes}
\end{threeparttable}
\end{table}
}
{\booksizes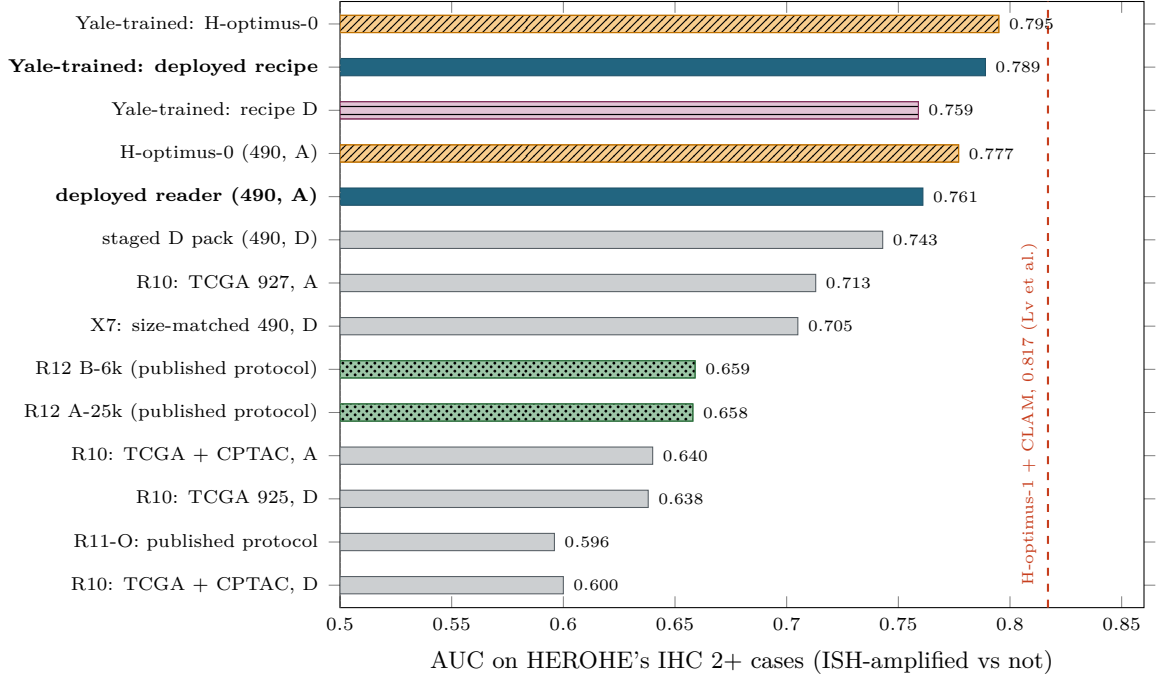
\begin{figure}[!htbp]
\centering
\begin{tikzpicture}
\begin{axis}[xbar, bar width=6.5pt, bar shift=0pt, width=0.74\textwidth, height=96mm, xmin=0.5, xmax=0.86,
  symbolic y coords={r10tcd,r11o,r10td,r10tca,r12a,r12b,x7,r10ta,dpack,dep,hopt,ydpack,ydep,yhopt}, ytick={r10tcd,r11o,r10td,r10tca,r12a,r12b,x7,r10ta,dpack,dep,hopt,ydpack,ydep,yhopt},
  yticklabels={{R10: TCGA + CPTAC, D}, {R11-O: published protocol}, {R10: TCGA 925, D}, {R10: TCGA + CPTAC, A}, {R12 A-25k (published protocol)}, {R12 B-6k (published protocol)}, {X7: size-matched 490, D}, {R10: TCGA 927, A}, {staged D pack (490, D)}, {\textbf{deployed reader (490, A)}}, {H-optimus-0 (490, A)}, {Yale-trained: recipe D}, {\textbf{Yale-trained: deployed recipe}}, {Yale-trained: H-optimus-0}},
  yticklabel style={font=\scriptsize}, xlabel={AUC on HEROHE's IHC 2+ cases (ISH-amplified vs not)}, xlabel style={font=\small},
  xticklabel style={font=\scriptsize}, nodes near coords, nodes near coords style={font=\tiny, /pgf/number format/.cd, fixed, zerofill, precision=3},
  enlarge y limits=0.04]
\addplot[fill=cgrey!35, draw=cgrey!80!black] coordinates {(0.600,r10tcd) (0.596,r11o) (0.638,r10td) (0.640,r10tca) (0.705,x7) (0.713,r10ta) (0.743,dpack)};
\addplot[fill=cok!50, draw=cok!80!black, postaction={pattern=crosshatch dots}] coordinates {(0.658,r12a) (0.659,r12b)};
\addplot[fill=cdata!75!black, draw=cdata!60!black] coordinates {(0.761,dep) (0.789,ydep)};
\addplot[fill=cgpu!45, draw=cgpu!80!black, postaction={pattern=north east lines}] coordinates {(0.777,hopt) (0.795,yhopt)};
\addplot[fill=cmodel!30, draw=cmodel!80!black, postaction={pattern=horizontal lines}] coordinates {(0.759,ydpack)};
\draw[cguard, thick, dashed] ({axis cs:0.817,r10tcd}|-{rel axis cs:0,0}) -- ({axis cs:0.817,r10tcd}|-{rel axis cs:0,1})
  node[pos=0.02, anchor=south west, font=\tiny, text=cguard, rotate=90] {H-optimus-1 + CLAM, 0.817 (Lv et al.)};
\draw[black!40, dotted] ({axis cs:0.5,r10tcd}|-{rel axis cs:0,0}) -- ({axis cs:0.5,r10tcd}|-{rel axis cs:0,1});
\end{axis}
\end{tikzpicture}
\caption{\emph{Measured} (HEROHE test split: 150 slides, 85 IHC 2+ of which 32 ISH-amplified; per-seed mean AUC of 10 or 8
seeds; readers trained on TCGA-BRCA unless marked Yale-trained). The hard cases: which IHC 2+ tumours are amplified. Solid
dark bars: the deployed reader's recipe. Dotted: R12's re-distilled encoders under the published protocol. Diagonal
hatching: H-optimus-0, the 1.1B teacher. Horizontal lines: the 1.0\,\textmu m/px recipe trained on Yale. Plain light bars:
every other training set and protocol we tried. Dashed line: the best published H\&E figure on HEROHE's IHC 2+ cases
(other protocol and training data~\cite{lv2026}). Chance is 0.5.}\label{fig:her2-hard}
\end{figure}}

\subsection{Training-set composition and per-laboratory levers}\label{sec:res-bias}

Our 490 TCGA slides favour large positives (\S\ref{sec:data}). Training on its 156 positives with negatives matched to
them in tissue area reads Yale at 0.81 instead of 0.85 at recipe D (0/10 seeds higher), so about 0.035--0.04 of Yale AUC
comes from that size cue; every representative training set we tried lands at 0.78--0.82 (Supplementary
Table~\ref{tab:xarms}). On HEROHE's equivocal cases, by contrast, our deployed reader's lead holds: the size-matched set
reads them at 0.71 and larger representative sets at 0.60--0.71, so most of the 490 slides' advantage there comes from
their positives and labels rather than tissue size.

\paragraph{Per-laboratory levers.} Two levers moved readers on a new site. Standardising each reader's scores over a
site's own unlabelled slides raised Yale balanced accuracy at the transferred threshold by 0.04--0.08 for readers on
Phikon, H-optimus-0 and Midnight-12k and halved the spread of sensitivity and specificity across seeds (0.10--0.20 to
0.04--0.10) for every reader, ours included, whose balanced accuracy barely changed (0.73 to 0.74). Training on TCGA-BRCA
plus that site's other validated slides, cross-validated so that each Yale slide is scored by models that never saw it,
raised Yale AUC from 0.81 to 0.83 for ours and by 0.06--0.07 for readers on H-optimus-0 and Phikon. A slide classifier
retrains in minutes with our encoder frozen, so both levers are open to any laboratory with its own cases.

\subsection{Against the best published reader}\label{sec:res-protocol}

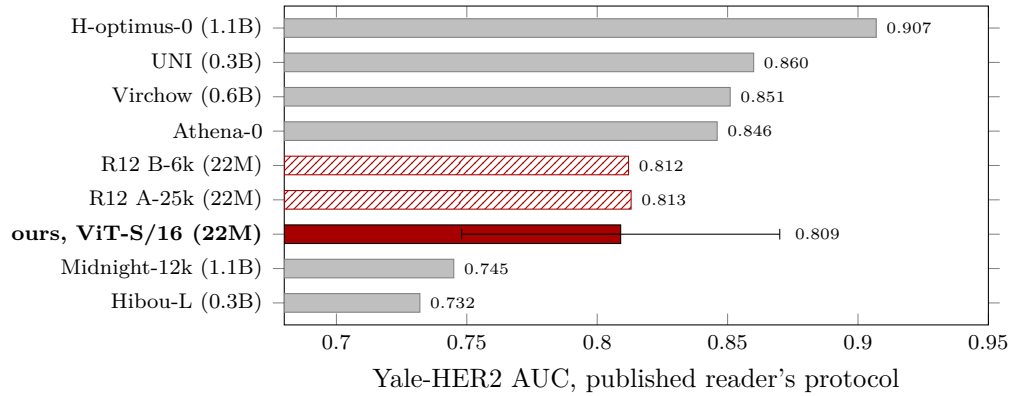
\begin{figure}[tbp]
\centering
\begin{tikzpicture}
\begin{axis}[xbar, bar width=7pt, bar shift=0pt, width=0.66\linewidth, height=58mm, xmin=0.68, xmax=0.95,
  symbolic y coords={hib, mid, ours, r12a, r12b, ath, vir, uni, hopt},
  ytick={hib, mid, ours, r12a, r12b, ath, vir, uni, hopt},
  yticklabels={{Hibou-L (0.3B)}, {Midnight-12k (1.1B)}, {\textbf{ours, ViT-S/16 (22M)}}, {R12 A-25k (22M)}, {R12 B-6k (22M)}, {Athena-0}, {Virchow (0.6B)}, {UNI (0.3B)}, {H-optimus-0 (1.1B)}},
  yticklabel style={font=\scriptsize}, xlabel={Yale-HER2 AUC, published reader's protocol}, xlabel style={font=\small},
  xticklabel style={font=\scriptsize}, nodes near coords, nodes near coords style={font=\tiny, /pgf/number format/.cd, fixed, zerofill, precision=3},
  enlarge y limits=0.08]
\addplot[fill=black!25, draw=black!50] coordinates {(0.732,hib) (0.745,mid) (0.846,ath) (0.851,vir) (0.860,uni) (0.907,hopt)};
\addplot[fill=red!65!black, draw=black, nodes near coords={}, error bars/.cd, x dir=both, x explicit] coordinates {(0.809,ours) +- (0.061,0)};
\addplot[draw=red!65!black, pattern=north east lines, pattern color=red!65!black] coordinates {(0.813,r12a) (0.812,r12b)};
\node[font=\tiny, anchor=west] at (axis cs:0.872,ours) {0.809};
\end{axis}
\end{tikzpicture}
\caption{HER2 status under the published reader's protocol: every tissue tile at 1.0\,\textmu m/px, attention MIL, a
40-model ensemble trained on TCGA-BRCA and CPTAC-BRCA, tested on Yale-HER2. Grey: as reported~\citep{athena2025}; dark:
our encoder (DeLong 95\% interval); hatched: R12's two re-distilled candidates; both on our replica of the protocol.}\label{fig:sota}
\end{figure}

With every tissue tile at 1.0\,\textmu m/px, non-gated attention MIL and a 40-model ensemble trained on TCGA-BRCA and
CPTAC-BRCA, our encoder reads Yale-HER2 at 0.81 (DeLong 95\% interval 0.75--0.87), close to UNI (0.86), Virchow (0.85)
and Athena-0 (0.85), above Midnight-12k and Hibou-L (0.75, 0.73), with H-optimus-0 at 0.91 as
published~\citep{athena2025} (Figure~\ref{fig:sota}). On HEROHE our default reader suits it better (0.75 on all 150
slides and 0.76 on IHC~2+, against 0.70 and 0.60 under this protocol). Distilling H-optimus-0 into our encoder again at
this protocol's reading scale (R12) gave two candidates at 0.813 and 0.812 on Yale, which fit training hospitals and
HEROHE's IHC~2+ cases better under this protocol (0.66 against 0.60) but fell short of the adoption bar fixed in advance
(0.829) and of our default reader on IHC~2+ (Supplementary~\ref{app:reader}).

\subsection{Which reader to deploy}\label{sec:res-review}

Table~\ref{tab:reader-review} sets every reader we built beside published state of the art. Our deployed reader
(five CLAM-SB classifiers on our final encoder, trained on our 490 TCGA slides with recipe A) is our best, or level with
our best, on every external hospital (Yale-HER2 0.82, BCNB 0.63, HEROHE 0.75) and on HEROHE's IHC~2+ cases (0.76). It
also costs least to run: 400 tiles, 5.2--5.6\,s per slide on a 12-core CPU, against an estimated 25\,s for the all-tile
protocol and about 5\,min for a reader on H-optimus-0. And it alone has been verified in service. We tested each
alternative against it under a rule fixed in advance, and none did better. We keep it as pathologists' baseline reader,
with three rules of use. Each laboratory sets its operating point, by site normalisation or on validated cases. On
IHC~2+ it orders and prioritises ISH without replacing it, and sends the uncertain middle to ISH or a second read
(Figure~\ref{fig:proto-zones}). And it is read beside IHC, never instead of it.

{\booksizes\begin{table}[!htbp]
\centering\scriptsize\setlength{\tabcolsep}{2.6pt}
\caption{\emph{Measured} unless marked. Every HER2 reader we built and tested, beside the published state of the art.}\label{tab:reader-review}
\begin{threeparttable}
\begin{tabularx}{\textwidth}{@{}>{\raggedright\arraybackslash}X>{\raggedright\arraybackslash}p{17mm}cccccc>{\raggedright\arraybackslash}p{11mm}@{}}
\toprule
reader & trained on & Yale & BCNB & HEROHE & HEROHE & inter- & Yale & read \\
 & & (192) & (1,058) & (150) & IHC 2+ & nal$^o$ & sens / spec & \\
\midrule
\multicolumn{9}{@{}l}{\emph{our readers: ViT-S/16 encoder (22M), CLAM-SB heads unless stated}} \\
\textbf{deployed} (her2-jepa2-t2y-s4): 400 tiles at 0.5\,\textmu m/px, 5 heads & TCGA 490 (size-biased) & \textbf{0.815}$^s$ & \textbf{0.628} & \textbf{0.746} & \textbf{0.761} & 0.687 & 0.72 / 0.79 & 5.2--5.6\,s \\
same recipe, trained on Yale & Yale 192 & --- & 0.701 & 0.836 & 0.789 & --- & --- & 5.2--5.6\,s \\
staged D pack: 400 tiles at 1.0\,\textmu m/px & TCGA 490 & 0.851 & 0.612 & 0.737 & 0.743 & --- & 0.91 / 0.45 & $\approx$6\,s$^e$ \\
R10: 400 tiles at 0.5\,\textmu m/px & TCGA 927 & 0.801 & --- & 0.702 & 0.713 & --- & --- & 5.2--5.6\,s \\
R11-O: all tiles at 1.0\,\textmu m/px, 40 ABMIL heads & TCGA 917 + CPTAC 382 & 0.809 & --- & 0.695 & 0.596 & 0.644 & 0.63 / 0.82 & $\approx$25\,s$^e$ \\
R12 A-25k (encoder re-distilled at 1.0\,\textmu m/px): as R11-O & TCGA 917 + CPTAC 382 & 0.813 & --- & 0.690 & 0.658 & 0.686 & 0.81 / 0.63 & $\approx$25\,s$^e$ \\
R12 B-6k (A-25k + JEPA): as R11-O & TCGA 917 + CPTAC 382 & 0.812 & --- & 0.718 & 0.659 & 0.689 & 0.83 / 0.61 & $\approx$25\,s$^e$ \\
\midrule
\multicolumn{9}{@{}l}{\emph{the teacher, same pipeline as the deployed reader (backup only)}} \\
H-optimus-0 (ViT-g, 1.1B): 400 tiles at 0.5\,\textmu m/px & TCGA 490 & 0.805 & 0.649 & 0.765 & 0.777 & --- & 0.41 / 0.89 & $\approx$5\,min \\
\midrule
\multicolumn{9}{@{}l}{\emph{published (Bosch et al.~\cite{athena2025}: all tiles at 1.0\,\textmu m/px, attention MIL, 40 models)}} \\
H-optimus-0 (1.1B) & TCGA 971 + CPTAC 323 & 0.907 & 0.716 & (508) & --- & --- & --- & GPU \\
UNI; Virchow; Athena-0 & TCGA + CPTAC & 0.86; 0.85; 0.85 & --- & --- & --- & --- & --- & GPU \\
\multicolumn{9}{@{}l}{\emph{published, other protocols}} \\
Virchow2 & --- & --- & --- & 0.855 & --- & --- & --- & GPU \\
H-optimus-1 + CLAM (Lv et al.~\cite{lv2026}) & --- & 0.900 & --- & --- & 0.817 & --- & --- & GPU \\
\bottomrule
\end{tabularx}
\begin{tablenotes}[flushleft]\scriptsize
\item[] AUC on hospitals a reader never trained on, per-seed mean unless marked. \emph{Read}: time per slide on the deployment server's 12-core CPU, archive to proposal. ---, not run. Published rows use their own protocols, training sets and test splits; their HEROHE is 508 slides, including its training split.
\item[$^s$] The shipped seed's five-model ensemble (10-seed mean 0.812).
\item[$^o$] The training cohort's out-of-fold AUC: TCGA 490 (one seed) for 400-tile readers, TCGA + CPTAC (8 seeds) for the published protocol.
\item[$^e$] Estimated from the tile count.
\end{tablenotes}
\end{threeparttable}
\end{table}
}

\subsection{Borderline HER2-low and FISH amplification}\label{sec:res-tasks}

\begin{table}[tbp]
\centering\footnotesize
\caption{Two finer HER2 questions on TCGA-BRCA.}\label{tab:tasks}
\begin{threeparttable}
\begin{tabularx}{\linewidth}{@{}>{\raggedright\arraybackslash}X r c cc@{}}
\toprule
 & & HER2-low vs HER2-zero & \multicolumn{2}{c}{FISH amplification, read by the HER2 reader} \\
\cmidrule(lr){3-3}\cmidrule(lr){4-5}
tile encoder & params & dedicated head & Yale-trained & TCGA-trained, out of fold \\
\midrule
Phikon & 86M & 0.741$^{1}$ & 0.619$^{6}$ & 0.654$^{3}$ \\
H-optimus-0 & 1.1B & 0.692$^{2}$ & 0.706$^{1}$ & 0.690$^{1}$ \\
Hibou-B & 86M & 0.678$^{3}$ & 0.672$^{3}$ & 0.572$^{5}$ \\
\textbf{Ours (final)} & 22M & \textbf{0.649$^{4}$} & \textbf{0.682$^{2}$} & \textbf{0.669$^{2}$} \\
Midnight-12k & 1.1B & 0.626$^{5}$ & 0.650$^{4}$ & 0.644$^{4}$ \\
ResNet-50$^\ast$ & 26M & 0.615$^{6}$ & 0.575$^{7}$ & 0.445$^{7}$ \\
DINOv2-S$^\ast$ & 22M & 0.613$^{7}$ & 0.632$^{5}$ & 0.469$^{6}$ \\
\bottomrule
\end{tabularx}

\begin{tablenotes}[flushleft]\footnotesize
\item[] AUC, mean over 10 paired seeds. Superscript: rank in its column. Standard deviations and paired tests in Supplementary Table~\ref{tab:tasks-full}.
\item[] \emph{HER2-low vs HER2-zero}: a slide classifier trained on that task's own labels (204 slides).
\item[] \emph{FISH amplification}: HER2-status readers used as they are, scoring \pp{t4-fish-n} FISH-tested slides by their HER2 probability without ever seeing a FISH label; trained on Yale-HER2, or trained on TCGA-BRCA and scored out of fold.
\item[$^\ast$] Natural-image controls.
\end{tablenotes}
\end{threeparttable}
\end{table}

\paragraph{HER2-low against HER2-zero.} This boundary now decides eligibility for trastuzumab
deruxtecan~\citep{modi2022}, and pathologists find it hard (\S\ref{sec:data}). It needs its own classifier: a status
reader counts both groups as ``negative'' (\pp{t4-lowx-min}--\pp{t4-lowo-max} across encoders; Supplementary
Table~\ref{tab:tasks-full}). Trained on this task's own labels (Table~\ref{tab:tasks}), Phikon reaches
\pp{t4-her2low-phikon}, H-optimus-0 \pp{t4-her2low-hopt}, and ours \pp{t4-her2low-ours}, ranking
\pp{t4-her2low-rank}th of \pp{t4-n}, ahead of Midnight-12k and both controls. Along our own training stages, distillation
lifts HER2-low to its teacher's level and JEPA lowers it again (Supplementary~\ref{app:tasks}).

\paragraph{FISH amplification from H\&E.} Can a reader see gene amplification in H\&E morphology? Without any FISH
label, readers built on our encoder rank amplified tumours above non-amplified ones at AUC \pp{t4-fishx-ours} when
trained on Yale-HER2 and \pp{t4-fisho-ours} when trained on TCGA-BRCA and scored out of fold, second of \pp{t4-n} both
times, after H-optimus-0 (\pp{t4-fishx-hopt}) and ahead of Phikon (\pp{t4-fishx-phikon}). Most of this signal is HER2
status itself: in this FISH-tested subset FISH agrees with recorded status for \pp{t4-fish-agree} of \pp{t4-fish-both}
slides, and a classifier trained on FISH labels alone, with only \pp{t4-fish-pos} amplified slides, stays near chance for
every encoder (\pp{t4-fish-min}--\pp{t4-fish-max}). So a reader can flag likely amplification before ISH without
replacing it. Equivocal 2+ tumours are too few here to test on their own; HEROHE tests them directly
(\S\ref{sec:res-hospitals}).

\subsection{CPU cost}\label{sec:res-cpu}

\begin{table}[tbp]
\centering\footnotesize
\caption{Running cost without a GPU.}\label{tab:cpu}
\begin{threeparttable}
\begin{tabularx}{\linewidth}{@{}>{\raggedright\arraybackslash}X l r r r r@{}}
\toprule
tile encoder & architecture & params & GMAC/tile & s/slide & slides/h \\
\midrule
\textbf{Ours (final)} & ViT-S/16 & 22M & \her{gmac-d16final} & \textbf{\her{cpu-d16final}} & \textbf{$\approx$430} \\
Phikon; Hibou-B & ViT-B/16; /14 & 86M & \her{gmac-phikon}; \her{gmac-hibou} & \her{cpu-phikon}$^a$ & $\approx$160 \\
H-optimus-0; Midnight-12k & ViT-g/14 & 1.1B & \her{gmac-hoptcard} & \her{cpu-hoptcard} & $\approx$12 \\
\midrule
\multicolumn{6}{@{}l}{\emph{Our shipped reader, deployed (node-optimised build, same server)}} \\
\quad per slide in service, archive to proposal & & & & 5.2--5.6 & $\approx$650 \\
\bottomrule
\end{tabularx}
\begin{tablenotes}[flushleft]\footnotesize
\item[] Tile encoder on a 12-core CPU server (32\,GB RAM), one library (numbat SDK 0.22.0), f32, one 400-tile slide. Slides per hour: one server, encoding only, sequential.
\item[$^a$] Timed on ViT-B/16; ViT-B/14 has 31\% more tokens per tile.
\end{tablenotes}
\end{threeparttable}
\end{table}

Model size shows up directly in running cost (Table~\ref{tab:cpu}). On GPUs every encoder here is fast. On a 12-core CPU
server, our ViT-S/16 encodes a 400-tile slide in \her{cpu-d16final}\,s, \her{cpux-phikon}$\times$ faster than Phikon's
ViT-B/16 (\her{cpu-phikon}\,s) and \her{cpux-hoptcard}$\times$ faster than H-optimus-0 (about 5\,min). Built for that
server, our shipped reader returns a HER2 proposal in under 6\,s per slide, about 650 slides an hour, and matches our
training pipeline to $6\times10^{-8}$. A small reader can serve a department from hardware it already owns, and on GPUs
it is cheaper too.

\section{Discussion}\label{sec:discussion}

\paragraph{A baseline for pathologists.} We call our deployed reader a baseline for two reasons. Of all readers we
built, it does best on hospitals it never saw (Yale-HER2 0.82, HEROHE 0.75, BCNB 0.63) and on equivocal IHC~2+ tumours
(0.76), at 5--6\,s per slide on a CPU, and we own every part of it. It also sets a level that a reader trained on a
laboratory's own data now has to beat. Under two independent protocols, our paired cross-cohort comparison and our
local PathBench replica, a reader on our 22M-parameter encoder sits with readers on Phikon (86M) and H-optimus-0 (1.1B),
and ranks first on TCGA-trained tests under both. A model a fiftieth of a
large foundation model's size can carry much of what that model knows about HER2 status, and it runs wherever a
department needs it.

\paragraph{Where it helps in the HER2 workflow.} Our reader cannot see protein, so it neither scores IHC nor grades
HER2-low. It helps before and beside IHC. It puts an early probability on every H\&E slide, flags a case whose confident
prediction disagrees with its IHC score, and ranks IHC~2+ tumours by likely amplification so that ISH requests can be
ordered. At AUC 0.76 on HEROHE's equivocal cases, close to the best published H\&E figure there (0.82, a larger encoder
under another protocol~\citep{lv2026}), that ranking sets priorities for ISH, which then decides each case. Two thresholds with an
abstain zone between them (Figure~\ref{fig:proto-zones}) send the uncertain middle to ISH or a second read. On new sites
its probabilities run low, so each laboratory should set its own operating point: site normalisation needs only
unlabelled local slides, and refitting on validated cases does better.

\paragraph{What moves a reader on a new site.} Laboratory-side levers move readers most. A site's own validated slides
in training raised Yale AUC by 0.02 for ours and 0.06--0.07 for larger encoders, a site-normalised threshold steadied
sensitivity and specificity across seeds, and readers trained on Yale-HER2 lead TCGA-trained ones by 0.06--0.10 on two
other hospitals. Since a slide classifier on our frozen encoder retrains in minutes, a laboratory can adapt our deployed
baseline to its own cases. On the encoder side, a larger distillation budget is the clearest next step: our re-distillation
at reading scale used 25,000 steps on 1.4M cached tiles, about 130 times less than its vendor's own
student~\citep{filiot2025}.

\paragraph{What a status reader knows beyond status.} Readers trained only on HER2 status rank FISH-amplified tumours
above non-amplified ones without having seen a FISH label, and readers on our encoder do so second best of seven. In our
cohort, though, FISH and recorded status mostly agree. This shows that a reader carries status information into
FISH-tested cases; it does not show that a reader measures gene copies. Equivocal 2+ tumours were too few in TCGA-BRCA to
test alone, and HEROHE's 85 IHC~2+ cases answer that question instead (\S\ref{sec:res-hospitals}). HER2-low against
HER2-zero behaves differently: a status reader treats both as negative and cannot separate them, and a classifier trained
for this boundary reaches 0.65--0.74 across encoders, with larger ones ahead. A reader that also proposes HER2-low needs
its own head for it and, with our encoder, more of what distillation gives (below).

\paragraph{Why JEPA helps.} Distillation gave our ViT-S most of H-optimus-0's representation and took it to
\mm{m5s25-ty} on external validation; changes to distillation itself (input matching, more data, longer schedules,
weight averaging) did not move it further (Supplementary Table~\ref{tab:tried}). JEPA asks our distilled encoder to
predict its own features of hidden patches from visible ones. On every ViT-S student distilled without student-side
augmentation it raised cross-cohort AUC (\TY{} +0.02 to +0.04, 9--10 of 10 seeds), consistent with a report that a
JEPA stage makes features hold up better across sites and scanners~\citep{genbio2026}. It did nothing for a student
already trained with student-side augmentation, and augmenting JEPA's own views changed nothing, which suggests that both
protect against similar shifts. ViT-B students gained less ($\approx$0.01).

\paragraph{Refreshing a target matters more than training longer.} A second round, with its target refreshed to round
one's result, added +0.015 to +0.020 on \TY{} across three independent runs, while one run of twice the length trailed
two rounds by 0.024 and a third round hurt. Refreshing its target is what helps, and two refreshes are enough here.
Updating that target continuously instead, as an exponential moving average of our encoder as in
I-JEPA~\citep{assran2023}, did not do better: across eight runs it raised \YY{} by up to 0.02 but cost 0.02--0.03 on
\TT, with \TY{} no higher. Two discrete refreshes keep a balance that a moving target does not.

\paragraph{Where JEPA starts matters.} From a mid-schedule distilled encoder (step 25,000), JEPA costs a little within a
cohort and gains most across cohorts. From that run's fully decayed end (step 100,000), it instead raises \YY{} by
\mm{pd-j100s-m5s-yy} to \mm{j100s-yy}, level with Phikon, while \TY{} rises less (\mm{j100s-ty}), and a second round
there gives back that \YY{} gain. So two starts suit two tests, and we could not combine them: interpolating their
weights lands between them, and JEPA with one start's encoder predicting the other's features falls below our final
encoder on every test. Blending our final encoder with its own distilled start works better, as weight-space ensembles
of a fine-tuned model and its start do elsewhere~\citep{wortsman2022}: at equal weights it keeps the start's \YY{}
(\mm{wisea50-yy}) and most of JEPA's \TY{} gain (\mm{wisea50-ty}), above a straight line between the two, but no blend
keeps \TY{} at our final encoder's level. Distillation at step 25,000 also holds HER2-low at its teacher's level
(\pp{t4s2-her2low}), and both JEPA and longer distillation erode it (Supplementary~\ref{app:tasks}).

\paragraph{Limitations.} Our 490 TCGA training slides favour large positives, worth about 0.04 of Yale AUC; we report
representative training sets beside them, and our reader's lead on equivocal cases holds under that correction. The
IHC~2+ evidence comes from one external set of 85 tumours, and Yale-HER2 (192 slides) and PathBench-style test splits are
small, so we rely on ten paired seeds and report paired tests. HEROHE and BCNB carry non-commercial licences and only test
our reader. Its operating point should be confirmed by each laboratory, and HER2-low grading needs a head of its own. Our
encoder inherits H-optimus-0's provenance: openly licensed weights trained on proprietary slides. Our reader is a
research tool that proposes a status for a pathologist to confirm, not a diagnostic device.

\section{Conclusion}\label{sec:conclusion}

We built and deployed a HER2 reader for routine H\&E slides, owned end to end, as a baseline for pathologists. Its
22M-parameter ViT-S/16 tile encoder was trained on public TCGA slides by self-supervision, distillation from H-optimus-0
and two rounds of JEPA refinement; five cross-validated CLAM slide classifiers are averaged, and an operating point is
fixed on training data. Among readers that differ only in their encoder it ranks first of \pp{rank-n} on external
validation (\TY{} \mm{jj25s3-ty}), alongside encoders 4 and 50 times its size. As a whole reader it reaches 0.82 on
Yale-HER2, 0.75 on HEROHE and 0.63 on BCNB, and among HEROHE's equivocal IHC~2+ cases it ranks ISH-amplified above
non-amplified tumours at 0.76, best among our TCGA-trained readers. A local replica of PathBench's benchmark,
validated against its published order of reference encoders, confirms our encoder's place beside Phikon and H-optimus-0
on public data and gives others a way to rank encoders for HER2 without private cohorts. At 5--6\,s per slide on a CPU, our reader serves
as a research-use pre-read in a pathology platform beside ER, PR and Ki-67. There it can order ISH for equivocal cases
and flag discordant ones, while IHC and the pathologist decide.

\paragraph{Next steps.} With a laboratory's own data: (i) retrain our slide classifiers on its validated cases and set its
operating point, two levers that moved readers on a new site; (ii) validate locally with IHC~2+ tumours and their ISH
results as the primary endpoint; (iii) run our replica of the published protocol with H-optimus-0 itself, to place our
encoder against the published 0.91 on one pipeline; and (iv) given a larger compute budget, distil the teacher at a scale
closer to its vendor's own student, every patch token included. HER2-low grading and IHC membrane scoring need heads of
their own.

\paragraph{Reproducibility and availability.} All tables and figures are regenerated by scripts from verbatim copies of
per-run result files, and every training run, reader experiment and departure from a published recipe was registered,
with its decision rule, before it ran. Model weights and output data are public at
\url{https://huggingface.co/cloudkites/her2-reader}: our ViT-S/16 encoder at every training stage, as timm-compatible
safetensors and as trained; the deployed reader (encoder, five slide classifiers, tile recipe and operating point); both
re-distilled candidates; per-slide predictions on TCGA-BRCA, CPTAC-BRCA and Yale-HER2 for every seed; and the data behind
every table and figure, with SHA-256 checksums. Weights carry CC~BY-NC~4.0 and data CC~BY~4.0. Per-slide results on HEROHE
and BCNB stay private under those cohorts' licences, so only aggregates appear here. Source code is not released; run
manifests, training logs and evaluation records are archived privately for audit.

\bibliographystyle{abbrvnat}
\bibliography{refs}

\clearpage
\appendix
\begin{center}{\Large\bfseries Supplementary material}\end{center}

\section{Full results under our protocol}\label{app:v23}

\emph{Our final encoder's numbers hold across independent runs, and every released encoder was scored in the same
way.} Table~\ref{tab:v23} gives mean and standard deviation over ten paired seeds for each released encoder
and for our ViT-S/16 after each stage. Three independent round-2 runs (seeds 3, 4 and 5 through both JEPA rounds) land
within 0.002 of one another on \TY{} and within 0.009 on every test.

\begin{table}[H]
\centering\footnotesize
\caption{HER2 status under our paired protocol, every encoder.}\label{tab:v23}
\begin{threeparttable}
\begin{tabularx}{\linewidth}{@{}>{\raggedright\arraybackslash}X cccc@{}}
\toprule
encoder & Yale$\to$Yale & TCGA$\to$TCGA & TCGA$\to$Yale & Yale$\to$TCGA \\
\midrule
\multicolumn{5}{@{}l}{\emph{Released encoders}} \\
\quad Phikon (ViT-B/16, 86M) & 0.872 (0.014) & 0.679 (0.012) & 0.811 (0.014) & 0.662 (0.003) \\
\quad H-optimus-0 (ViT-g/14, 1.1B) & 0.897 (0.012) & 0.676 (0.017) & 0.805 (0.020) & 0.653 (0.010) \\
\quad Hibou-B (ViT-B/14, 86M) & 0.786 (0.025) & 0.626 (0.011) & 0.749 (0.027) & 0.637 (0.011) \\
\quad Midnight-12k (ViT-g/14, 1.1B) & 0.834 (0.018) & 0.623 (0.019) & 0.727 (0.029) & 0.638 (0.005) \\
\quad ResNet-50, ImageNet (26M) & 0.700 (0.022) & 0.545 (0.017) & 0.638 (0.022) & 0.568 (0.015) \\
\quad DINOv2 ViT-S/14, natural images (22M) & 0.676 (0.029) & 0.495 (0.016) & 0.589 (0.016) & 0.573 (0.009) \\
\multicolumn{5}{@{}l}{\emph{Our ViT-S/16 after each stage}} \\
\quad Stage 1: self-supervised & 0.761 (0.027) & 0.589 (0.020) & 0.589 (0.047) & 0.520 (0.006) \\
\quad Stage 2: distilled (step 25,000) & 0.858 (0.022) & 0.705 (0.009) & 0.762 (0.011) & 0.625 (0.007) \\
\quad Stage 3, JEPA round 1, seed 3 & 0.848 (0.015) & 0.689 (0.014) & 0.798 (0.011) & 0.642 (0.009) \\
\quad Stage 3, JEPA round 1, seed 4 & 0.851 (0.014) & 0.687 (0.019) & 0.792 (0.016) & 0.642 (0.008) \\
\quad Stage 3, JEPA round 1, seed 5 & 0.850 (0.012) & 0.687 (0.018) & 0.795 (0.014) & 0.643 (0.008) \\
\quad Stage 3, JEPA round 2 from seed 3 (final) & 0.844 (0.022) & 0.690 (0.012) & 0.812 (0.015) & 0.650 (0.008) \\
\quad Stage 3, JEPA round 2 from seed 4 & 0.847 (0.018) & 0.681 (0.016) & 0.812 (0.010) & 0.651 (0.006) \\
\quad Stage 3, JEPA round 2 from seed 5 & 0.853 (0.019) & 0.685 (0.013) & 0.810 (0.015) & 0.652 (0.007) \\
\bottomrule
\end{tabularx}

\begin{tablenotes}[flushleft]\footnotesize
\item[] AUC, mean (SD) over 10 paired seeds; one software build (numbat SDK 0.23.0).
\end{tablenotes}
\end{threeparttable}
\end{table}

\section{Ranking on a local PathBench replica}\label{app:pb}

\emph{Our 22M-parameter encoder ranks first of 14 encoder readouts on TCGA-BRCA and sixth on Yale-HER2, and no
foundation model is ahead of it at $p<0.05$ on either cohort.} PathBench evaluates encoders on private cohorts held out of
all pretraining~\citep{pathbench2025}, so its numbers cannot be reproduced on public data. Our replica of its
classification protocol (\S\ref{sec:method-eval}) trains an attention-based MIL classifier (ABMIL) from scratch for each
task, encoder and seed, on one case-stratified 7:1:2 train/validation/test split per seed, seeds 1--10, shared by every
encoder (TCGA-BRCA: 343/49/98 slides; Yale-HER2: 38 test slides); it reports test AUC with a 1,000-sample bootstrap and
exact one-sided paired Wilcoxon tests. It passes its check: on both cohorts it orders PathBench's reference encoders on
breast HER2 exactly as PathBench does (H-optimus-0 0.845 $>$ Phikon 0.814 $>$ Hibou 0.780 $>$ ResNet-50
0.715)~\citep{pathbench2025}. Figure~\ref{fig:pb-rank} ranks our final encoder among every pathology foundation model
and control we scored this way, with H-optimus-0 and Midnight-12k under each readout we tried. Main-text
Table~\ref{tab:pb-rank} keeps one row per model. On TCGA-BRCA our encoder leads H-optimus-0 in 6 to 9 of 10 paired seeds
under each of its four readouts, and is ahead of Midnight-12k, Hibou-B and both controls ($p\leq0.007$). On Yale-HER2
H-optimus-0's readouts and Phikon lead it by 0.004 to 0.016 ($p\geq0.08$ for each), and it is ahead of Hibou-B, our own
self-supervised stage and both controls ($p\leq0.01$). We extracted our features with numbat SDK 0.23.0 and others' with
0.15.3, which differ by at most $1.2\times10^{-4}$.

\begin{figure}[H]
\centering
\begin{tikzpicture}
\begin{axis}[hr panel, table/col sep=tab, scale only axis, width=52mm, height=14*3.4mm, y dir=reverse, ymin=0.4,
  ymax=14.6, xmin=0.35, xmax=0.82, ytick=data, yticklabels from table={data/pb_rank_tcga.dat}{label},
  yticklabel style={font=\scriptsize}, major y tick style={draw=none}, xmajorgrids,
  xlabel={test AUC, mean of 10 seeds}, title={\textbf{(a)} TCGA-BRCA (98 test slides per seed)},
  xticklabel style={/pgf/number format/fixed, /pgf/number format/precision=2}]
\addplot[draw=none, mark=none, forget plot] table[x expr=0.35, y=y] {data/pb_rank_tcga.dat};
\addplot[black!45, only marks, mark=none, error bars/.cd, x dir=both, x explicit, error mark=|,
  error mark options={mark size=1pt}]
  table[x=mean, y=y, x error plus expr=\thisrow{hi}-\thisrow{mean}, x error minus expr=\thisrow{mean}-\thisrow{lo}]
  {data/pb_rank_tcga.dat};
\addplot[hr groups, only marks] table[x=mean, y=y, meta=group] {data/pb_rank_tcga.dat};
\end{axis}
\end{tikzpicture}\par\medskip
\begin{tikzpicture}
\begin{axis}[hr panel, table/col sep=tab, scale only axis, width=52mm, height=14*3.4mm, y dir=reverse, ymin=0.4,
  ymax=14.6, xmin=0.45, xmax=1.00, ytick=data, yticklabels from table={data/pb_rank_yale.dat}{label},
  yticklabel style={font=\scriptsize}, major y tick style={draw=none}, xmajorgrids,
  xlabel={test AUC, mean of 10 seeds}, title={\textbf{(b)} Yale-HER2 (38 test slides per seed)},
  xticklabel style={/pgf/number format/fixed, /pgf/number format/precision=2}]
\addplot[draw=none, mark=none, forget plot] table[x expr=0.45, y=y] {data/pb_rank_yale.dat};
\addplot[black!45, only marks, mark=none, error bars/.cd, x dir=both, x explicit, error mark=|,
  error mark options={mark size=1pt}]
  table[x=mean, y=y, x error plus expr=\thisrow{hi}-\thisrow{mean}, x error minus expr=\thisrow{mean}-\thisrow{lo}]
  {data/pb_rank_yale.dat};
\addplot[hr groups, only marks] table[x=mean, y=y, meta=group] {data/pb_rank_yale.dat};
\end{axis}
\end{tikzpicture}
\caption{HER2 status on our local PathBench replica: encoders ranked by mean test AUC over 10 paired seeds (one split per
seed, shared by all encoders); whisker, mean of the per-seed bootstrap 95\% intervals. Star: our final encoder;
triangles: pathology foundation models, H-optimus-0 and Midnight-12k under each readout we scored; circles: our
self-supervised stage alone and natural-image or untrained controls. In parentheses, PathBench's published breast-HER2 AUC
for its four reference encoders, whose order our replica reproduces (PathBench ranked Hibou-L; we hold Hibou-B).}
\label{fig:pb-rank}
\end{figure}

\section{Borderline and FISH tasks in full}\label{app:tasks}

\emph{A HER2-status reader carries amplification signal; HER2-low needs a classifier of its own, where larger
encoders lead.} Both tasks use TCGA-BRCA's 400-tile bags,
with labels from GDC's clinical file (Section~\ref{sec:data}); Yale-HER2 has no such labels, so neither task has external
validation. We read each task in three ways (Table~\ref{tab:tasks-full}): a dedicated CLAM-SB classifier trained on the
task's own labels (5-fold case-stratified cross-validation, 10 paired seeds, pooled out-of-fold AUC), and a HER2-status
reader used as it is, trained either on Yale-HER2 (applied to TCGA-BRCA, never seen) or on TCGA-BRCA (scored out of
fold). A dedicated FISH classifier sees only \pp{t4-fish-pos} amplified slides, too few to train from scratch; that is
why every encoder stays near chance there while status readers do far better.

\begin{table}[H]
\centering\scriptsize
\caption{Borderline and FISH tasks on TCGA-BRCA, every reading.}\label{tab:tasks-full}
\begin{threeparttable}
\begin{tabularx}{\linewidth}{@{}>{\raggedright\arraybackslash}X c c c c c c c@{}}
\toprule
test & \textbf{ours} & Phikon & H-optimus-0 & Hibou-B & Midnight-12k & ResNet-50 & DINOv2-S \\
\midrule
HER2-low, head & \textbf{0.649$^{4}$} & 0.741$^{1}$ & 0.692$^{2}$ & 0.678$^{3}$ & 0.626$^{5}$ & 0.615$^{6}$ & 0.613$^{7}$ \\
\quad SD; $p$ vs ours & 0.024; -- & 0.028; $<$0.01 & 0.026; $<$0.01 & 0.014; 0.02 & 0.036; 0.08 & 0.031; 0.02 & 0.037; $<$0.01 \\
HER2-low, reader Y & \textbf{0.456$^{6}$} & 0.479$^{4}$ & 0.500$^{2}$ & 0.542$^{1}$ & 0.494$^{3}$ & 0.462$^{5}$ & 0.441$^{7}$ \\
\quad SD; $p$ vs ours & 0.010; -- & 0.013; $<$0.01 & 0.007; $<$0.01 & 0.015; $<$0.01 & 0.010; $<$0.01 & 0.014; 0.28 & 0.019; 0.03 \\
HER2-low, reader T & \textbf{0.560$^{4}$} & 0.551$^{6}$ & 0.569$^{2}$ & 0.553$^{5}$ & 0.570$^{1}$ & 0.541$^{7}$ & 0.565$^{3}$ \\
\quad SD; $p$ vs ours & 0.026; -- & 0.044; 0.28 & 0.030; 0.28 & 0.032; 0.31 & 0.029; 0.28 & 0.028; 0.05 & 0.051; 0.31 \\
FISH, head & \textbf{0.544$^{2}$} & 0.475$^{5}$ & 0.546$^{1}$ & 0.485$^{4}$ & 0.538$^{3}$ & 0.468$^{7}$ & 0.471$^{6}$ \\
\quad SD; $p$ vs ours & 0.022; -- & 0.017; $<$0.01 & 0.025; 0.35 & 0.029; $<$0.01 & 0.034; 0.28 & 0.022; $<$0.01 & 0.048; $<$0.01 \\
FISH, reader Y & \textbf{0.682$^{2}$} & 0.619$^{6}$ & 0.706$^{1}$ & 0.672$^{3}$ & 0.650$^{4}$ & 0.575$^{7}$ & 0.632$^{5}$ \\
\quad SD; $p$ vs ours & 0.019; -- & 0.007; $<$0.01 & 0.009; $<$0.01 & 0.015; 0.12 & 0.005; $<$0.01 & 0.016; $<$0.01 & 0.014; $<$0.01 \\
FISH, reader T & \textbf{0.669$^{2}$} & 0.654$^{3}$ & 0.690$^{1}$ & 0.572$^{5}$ & 0.644$^{4}$ & 0.445$^{7}$ & 0.469$^{6}$ \\
\quad SD; $p$ vs ours & 0.024; -- & 0.020; 0.04 & 0.034; 0.04 & 0.034; $<$0.01 & 0.030; 0.01 & 0.030; $<$0.01 & 0.035; $<$0.01 \\
\bottomrule
\end{tabularx}

\begin{tablenotes}[flushleft]\scriptsize
\item[] AUC, mean over 10 paired seeds (superscript: rank in its row), then standard deviation and exact one-sided Wilcoxon $p$ that the higher of each pair is ahead of ours.
\item[] \emph{head}: dedicated classifier; \emph{reader Y}: HER2-status reader trained on Yale-HER2; \emph{reader T}: trained on TCGA-BRCA, scored out of fold.
\end{tablenotes}
\end{threeparttable}
\end{table}

Along our own training stages (Table~\ref{tab:ladder-low}), distillation lifts HER2-low to its teacher's level by step
25,000 (\pp{t4s2-her2low}, against H-optimus-0's \pp{t4-her2low-hopt}); both JEPA rounds and continued distillation
then lower it. So distilling H-optimus-0 alone cannot reach Phikon on this task, and future JEPA variants should aim to
keep what distillation gives.

\begin{table}[H]
\centering\footnotesize
\caption{HER2-low against HER2-zero along our training stages.}\label{tab:ladder-low}
\begin{threeparttable}
\begin{tabularx}{\linewidth}{@{}>{\raggedright\arraybackslash}X c c@{}}
\toprule
our ViT-S/16 & AUC (SD) & against stage 2 \\
\midrule
stage 1: self-supervised & 0.662 (0.015) & $-$0.020 (8/10 lower) \\
stage 2: distilled, step 25,000 & \textbf{0.683} (0.014) & -- \\
\quad + JEPA round 1 & 0.661 (0.029) & $-$0.022 (7/10 lower) \\
\quad + JEPA round 2 (\textbf{final}) & 0.649 (0.024) & $-$0.034 (8/10 lower) \\
\quad 0.75 final + 0.25 stage 2 (weights) & 0.651 (0.029) & $-$0.031 (7/10 lower) \\
\quad 0.5 final + 0.5 stage 2 (weights) & 0.665 (0.028) & $-$0.018 (6/10 lower) \\
distillation continued to step 100,000 & 0.649 (0.024) & $-$0.034 (10/10 lower) \\
\quad + JEPA & 0.629 (0.030) & $-$0.054 (10/10 lower) \\
\midrule
H-optimus-0 (teacher) / Phikon & 0.692 / 0.741 & \\
\bottomrule
\end{tabularx}

\begin{tablenotes}[flushleft]\footnotesize
\item[] AUC, mean (SD) over 10 paired seeds, and paired change against stage 2 (seeds of 10 on which a row is lower).
\end{tablenotes}
\end{threeparttable}
\end{table}

\section{Our reader programme in detail}\label{app:reader}

\emph{Our deployed reader's recipe held against every alternative on external hospitals, and size bias in its training
set explains why its Yale level sits above that of every representative set.} Each experiment below was registered with
its decision rule before it ran; readers are CLAM-SB with settings from \S\ref{sec:method-head}, ten seeds, mean
per-seed AUC.

\begin{table}[H]
\centering\footnotesize
\caption{Tile recipes for our encoder, tested on Yale-HER2.}\label{tab:recipes}
\begin{threeparttable}
\begin{tabularx}{\linewidth}{@{}l>{\raggedright\arraybackslash}Xcccc@{}}
\toprule
recipe & tiles & AUC (SD) & sens / spec & bal.\ acc. & PPV / NPV \\
\midrule
A$'$ (default, rebuilt) & $\le$400 of 128\,\textmu m at 0.5\,\textmu m/px & 0.805 (0.020) & 0.69 / 0.77 & 0.730 & 0.36 / 0.93 \\
C & $\le$2,000 of 128\,\textmu m at 0.5\,\textmu m/px & 0.806 (0.009) & 0.68 / 0.79 & 0.734 & 0.37 / 0.93 \\
B & $\le$2,000 of 224\,\textmu m at 1.0\,\textmu m/px & 0.833 (0.013) & 0.70 / 0.79 & 0.745 & 0.38 / 0.94 \\
D & $\le$400 of 224\,\textmu m at 1.0\,\textmu m/px & 0.851 (0.016) & 0.69 / 0.84 & 0.765 & 0.47 / 0.94 \\
\bottomrule
\end{tabularx}
\begin{tablenotes}[flushleft]\footnotesize
\item[] Readers trained on the 490 TCGA-BRCA slides; one software build. Sensitivity and specificity after site normalisation; PPV and NPV at 15\% prevalence.
\end{tablenotes}
\end{threeparttable}
\end{table}

On BCNB (1,058 biopsies) recipe D read 0.612 and 0.672 against A's 0.628 and 0.701 (TCGA- and Yale-trained), and on HEROHE
0.737 and 0.806 against 0.746 and 0.836, so we kept A. Neither a slide classifier without CLAM's instance branch (gated
or not; Yale $\pm$0.006) nor CLAM's multi-branch variant ($-0.04$ to $-0.07$) did better, and guideline-cleaned TCGA labels
lowered our reader's Yale AUC by 0.008 over 20 seeds while raising H-optimus-0's by 0.044.

\begin{table}[H]
\centering\footnotesize
\caption{Every training set we tried for our reader, tested on Yale-HER2 and HEROHE.}\label{tab:xarms}
\begin{threeparttable}
\begin{tabularx}{\linewidth}{@{}l>{\raggedright\arraybackslash}Xrrccc@{}}
\toprule
arm & training set & slides & positive & Yale & HEROHE & IHC~2+ \\
\midrule
control & the 490 (size-biased) & 488 & 32\% & 0.847 & 0.737 & 0.743 \\
X1 & 904 without the added IHC~2+ negatives & 764 & 25\% & 0.762 & --- & --- \\
X2 & 904, additions labelled like the 490 only & 731 & --- & 0.817 & --- & --- \\
X3 & 904, each site's class mix balanced & 536 & 36\% & 0.757 & --- & --- \\
X4 & 904 at 50\% positive & 386 & 50\% & 0.729 & --- & --- \\
X5 & 904 without the 37 added positives & 867 & 18\% & 0.813 & --- & --- \\
X6 & 904 at 10\% positive & 790 & 10\% & 0.804 & --- & --- \\
X7 & the 490's positives, size-matched negatives & 483 & 32\% & 0.810 & 0.721 & 0.705 \\
\midrule
R10 & TCGA 925 & 925 & 21\% & 0.790 & 0.698 & 0.638 \\
R10 & TCGA 925 + CPTAC-BRCA 382 & 1,307 & 18\% & 0.794 & 0.684 & 0.600 \\
R10$^a$ & TCGA 927 & 927 & 21\% & 0.801 & 0.702 & 0.713 \\
R10$^a$ & TCGA 927 + CPTAC-BRCA 388 & 1,315 & 18\% & 0.780 & 0.689 & 0.640 \\
\bottomrule
\end{tabularx}
\begin{tablenotes}[flushleft]\footnotesize
\item[] TCGA-trained readers at recipe D unless marked, against the 490-slide control. HEROHE: all 150 test slides and its 85 IHC~2+ cases. X1--X7: registered diagnostics of the 904 TCGA slides first fetched. ---, not measured.
\item[$^a$] Recipe A, against its own control (Yale 0.818).
\end{tablenotes}
\end{threeparttable}
\end{table}

\paragraph{Per-laboratory levers.} \emph{Site normalisation}: $z = (\mathrm{logit}\,p - \mu)/\sigma$ over each site's own
slides (the training cohort's out-of-fold predictions; the new site's unlabelled slides), with the Youden threshold set in
$z$ on the training cohort. Yale balanced accuracy, raw to normalised: ours 0.732 to 0.736, H-optimus-0 0.648 to 0.710,
Phikon 0.689 to 0.732, Hibou-B $+0.043$, Midnight-12k $+0.082$; the standard deviation of sensitivity and specificity
across seeds fell from 0.10--0.20 to 0.04--0.10. Calibrated probabilities did not transfer (expected calibration error on
Yale 0.25--0.40), and a rule-out threshold set at 95\% TCGA sensitivity gave Yale specificity near zero.
\emph{A site's own slides}: TCGA-BRCA (490) and Yale-HER2 (192) pooled into one cross-validation, each Yale slide scored by
models that saw TCGA and the other Yale folds: Yale AUC 0.834 for ours (TCGA only 0.812; Yale only 0.844), 0.866 for
H-optimus-0 and 0.883 for Phikon.

\paragraph{The published protocol and R12.} Our encoder's 40-model reader under the published protocol read Yale at
0.809 [0.748, 0.870] (single seeds 0.797, SD 0.023; sensitivity / specificity 0.63 / 0.82 at the training threshold), the
training cohort's out-of-fold AUC 0.644, HEROHE 0.695 (ensemble 0.707) and IHC~2+ 0.596. R12's candidates: A-25k Yale
0.813 [0.751, 0.875], 4 of 8 seeds above ours; B-6k 0.812 [0.751, 0.873], 5 of 8; out-of-fold 0.686 and 0.689; HEROHE 0.690
and 0.718, IHC~2+ 0.658 and 0.659. R12's tiles: 1,082,525 at 1.0\,\textmu m/px from 2,501 TCGA slides (TCGA-BRCA and 31 other projects) and 388
CPTAC-BRCA series; teacher cache 1.4M records (1.0M at 1.0\,\textmu m/px, 0.4M at 0.5), every rank's first chunk within
$2.3\times10^{-3}$ of an f32 recompute; distillation 25,000 steps (batch 64) from our final encoder, then one JEPA round of
6,000 steps.

\section{What else we tried}\label{app:tried}

\emph{JEPA refinement, applied twice from a mid-schedule distilled encoder, gave our largest external-validation gain;
our plain recipe matched or beat every variant we tried.}
Table~\ref{tab:tried} lists every change we tested, each registered before it ran and scored under our ten-seed
protocol as a paired difference. We compare distillation changes with the distilled student they modify, and JEPA
variants with plain JEPA from an identical start, unless noted.

\begin{table}[H]
\centering\scriptsize
\caption{Changes we tested and their effect.}\label{tab:tried}
\begin{threeparttable}
\begin{tabularx}{\linewidth}{@{}>{\raggedright\arraybackslash}p{50mm}>{\raggedright\arraybackslash}X>{\raggedright\arraybackslash}p{27mm}@{}}
\toprule
change & effect & verdict \\
\midrule
\multicolumn{3}{@{}l}{\emph{Distillation (stage 2)}} \\
evaluation input matched to training input (scale, crop, JPEG) & student \TY{} +0.009, Phikon lower; ranking unchanged & not the cause \\
tissue mix of training stream & 7.9--9.9\% breast tiles in every cache & not the cause \\
no patch-token targets & ViT-B \TY{} \mm{pd-k1b-m5b25-ty}; ViT-S \YT{} \mm{pd-k1s-m5s25-yt} & keep patch targets \\
photometric jitter of teacher's view & crop-only view better: \YY{} \mm{pd-m6ec-m6ef-yy}, \YT{} \mm{pd-m6ec-m6ef-yt} & avoid \\
2.5$\times$ more distillation data & \TY{} \mm{pd-v5s-v4s-ty} & no gain \\
longer distillation (ViT-B, 75k vs 50k steps) & \TY{} \mm{pd-m5b75-m5b50-ty} & hurts \\
ViT-B/16 student instead of ViT-S & best \TY{} \mm{m5b50-ty} & no better \\
weight averaging of runs or snapshots & at members' mean on \TY, or trades \YY{} for it & no gain \\
\multicolumn{3}{@{}l}{\emph{JEPA refinement (stage 3)}} \\
\textbf{JEPA, 6,000 steps, LR $10^{-4}$, from step 25,000} & \TY{} +0.030 to +0.036 (3 seeds), \YY{} $-$0.006 to $-$0.010 & \textbf{kept} \\
\textbf{second round (refreshed target)} & \TY{} +0.015 to +0.020 more (3 chains), \YY{} within 0.004 & \textbf{kept} \\
third round & \TY{} \mm{pd-jjj25s3-jj25s3-ty}, \YY{} \mm{pd-jjj25s3-jj25s3-yy} & stop at two \\
one run of 12,000 steps instead of two rounds & \TY{} \mm{pd-j2512k-jj25s3-ty} against two rounds & refresh, not length \\
learning rate $3\times10^{-5}$ / $2\times10^{-4}$ & smaller \TY{} gain (\mm{pd-jlr3-m6ds-ty} vs \mm{pd-jms3-m6ds-ty}; \mm{pd-j25lr2-j25s3-ty}) & keep $10^{-4}$ \\
2,000 decayed steps & \YY{} cost doubles (\mm{pd-jsh3-m6ds-yy} vs \mm{pd-jms3-m6ds-yy}) & no \\
flip and stain augmentation & \TY{} \mm{pd-j25aug-j25s3-ty} & no gain \\
block masks & \TY{} \mm{pd-j25blk-j25s3-ty} & no gain \\
start from distilled step 100,000 & \YY{} \mm{pd-j100s-m5s-yy} (to \mm{j100s-yy}, level with Phikon), \TY{} \mm{j100s-ty} & complementary start \\
second round from step 100,000 & \YY{} \mm{pd-jj100s3-j100s-yy}, \TY{} \mm{pd-jj100s3-j100s-ty} & no \\
student trained with student-side augmentation & \TY{} \mm{pd-jaug25-k4s-ty} & no gain \\
ViT-B starts & \TY{} +0.007 to +0.011 & smaller gain \\
distilling again after JEPA & \TY{} \mm{pd-rda-jms3-ty} & erases JEPA's gain \\
target updated continuously (EMA; 8 runs, $M$ 0.999 / 0.9995, 6,000 / 12,000 steps) & against two rounds: \YY{} +0.006 to +0.021, \TT{} $-$0.020 to $-$0.034, \TY{} 0 to $-$0.013 & two refreshes better \\
weight interpolation of final encoder and JEPA from step 100,000 ($\alpha$ 0.75, 0.5) & \TY{} \mm{pd-mixa75-jj25s3-ty}, \mm{pd-mixa50-jj25s3-ty}; \YY{} \mm{pd-mixa75-jj25s3-yy}, \mm{pd-mixa50-jj25s3-yy} & lands between members \\
cross-target JEPA (one start predicts the other's features) & \TY{} \mm{pd-xtf-jj25s3-ty}, \mm{pd-xth-jj25s3-ty}; \TT{} \mm{pd-xtf-jj25s3-tt}, \mm{pd-xth-jj25s3-tt} & no \\
weight interpolation of final encoder and its distilled start ($\alpha$ 0.75, 0.5) & \YY{} \mm{pd-wisea75-jj25s3-yy}, \mm{pd-wisea50-jj25s3-yy}; \TY{} \mm{pd-wisea75-jj25s3-ty}, \mm{pd-wisea50-jj25s3-ty} & better than a straight line; trades \TY{} for \YY \\
\bottomrule
\end{tabularx}
\begin{tablenotes}[flushleft]\scriptsize
\item[] Paired difference over 10 seeds on external validation (\TY) and within Yale (\YY). Bold: our kept recipe.
\end{tablenotes}
\end{threeparttable}
\end{table}

\section{Training details}\label{app:details}

\emph{Every stage fits on one laboratory's consumer GPUs, and every component was verified before use.}

\begin{table}[H]
\centering\footnotesize
\caption{Training recipe of our final encoder.}\label{tab:recipe}
\begin{threeparttable}
\begin{tabularx}{\linewidth}{@{}>{\raggedright\arraybackslash}p{24mm}>{\raggedright\arraybackslash}X>{\raggedright\arraybackslash}X>{\raggedright\arraybackslash}X@{}}
\toprule
 & stage 1: self-supervised & stage 2: distillation & stage 3: JEPA (twice) \\
\midrule
objective & DINO + iBOT + KoLeo, EMA teacher & $1-\cos$ to H-optimus-0's CLS token and $4\times4$ patch grid & mean squared error on own features of hidden patches \\
data & 98.6M TCGA tiles, one pass & 2M-tile cache, centre view & 2.1M TCGA tiles per round \\
steps $\times$ batch & 93,338 $\times$ 1,056 & 25,000 $\times$ 64 (of a 100,000-step schedule) & 6,000 $\times$ 352 per round \\
optimiser & AdamW, peak $1.0\times10^{-3}$, cosine & AdamW, $2\times10^{-4}$, cosine & AdamW, $10^{-4}$, cosine \\
masking, views & 10--50\% patch masks; stain jitter & stain jitter $\sigma=0.05$ & 20--50\% patches hidden; no augmentation \\
hardware & 3 RTX 3090 & 1 RTX 3090, $\approx$2.5\,h & 1 RTX 3090, $\approx$1\,h per round \\
\bottomrule
\end{tabularx}
\begin{tablenotes}[flushleft]\footnotesize
\item[] Each departure from a published recipe was registered in our design record before use.
\end{tablenotes}
\end{threeparttable}
\end{table}

\paragraph{HER2 classifier.} CLAM-SB with a 512-unit projection, gated attention of width 384, a linear classifier and an
instance-clustering branch used in training only; Adam $2\times10^{-4}$, weight decay $10^{-5}$, dropout 0.25, at most
200 epochs with patience 20, 5-fold case-stratified cross-validation; 0.59M parameters per model on ViT-S features.

\paragraph{Stain augmentation instead of colour jitter.}\label{app:aug} With DINOv2's photometric stack, two views of an
H\&E tile were nearly as dissimilar as views of different tiles: a training-free retrieval test found a tile's second
view among all second views in 0.188 of cases, against 0.063 by chance, and self-distillation collapsed to a uniform
prototype assignment. Bisecting our implementation pointed to this augmentation. Stain-space jitter (deconvolution,
per-concentration affine noise $\sigma=0.05$, reconstruction) keeps views related and costs 3.5\% of the student's
confidence.

\paragraph{Verification.}\label{app:impl} All code is our own, on numbat~\citep{tran2026numbat}. Before any long run,
each component passed numeric-gradient, CPU$\leftrightarrow$CUDA, overfit and resume gates (a run killed mid-schedule
ends byte-identical, including JEPA's moving target). Forward parity with PyTorch is cosine $\geq0.9999999$ for our ViT-S
and ViT-B; both distillation teachers match their PyTorch references to $3\times10^{-6}$.

\begin{table}[H]
\centering\footnotesize
\caption{Backbones compared for cost.}\label{tab:backbones}
\begin{threeparttable}
\begin{tabularx}{\linewidth}{@{}>{\raggedright\arraybackslash}p{30mm}>{\centering\arraybackslash}X>{\centering\arraybackslash}X>{\centering\arraybackslash}X@{}}
\toprule
 & ViT-S/16 (ours) & ViT-B/16 (Phikon) & ViT-g/14 (H-optimus-0) \\
\midrule
patch; tokens per tile & 16\,px; 196 + CLS & 16\,px; 196 + CLS & 14\,px; 256 + CLS + 4 reg. \\
width / heads / blocks & 384 / 6 / 12 & 768 / 12 / 12 & 1,536 / 24 / 40 \\
parameters & \her{par-d16final} & 86M & \her{par-hoptcard} \\
MACs per tile & \her{gmac-d16final}\,G & \her{gmac-phikon}\,G & \her{gmac-hoptcard}\,G \\
CPU s per 400-tile slide & \her{cpu-d16final} & \her{cpu-phikon} & \her{cpu-hoptcard} \\
\bottomrule
\end{tabularx}
\begin{tablenotes}[flushleft]\footnotesize
\item[] MACs computed from each architecture; CPU time measured (Supplementary~\ref{app:cpu}).
\end{tablenotes}
\end{threeparttable}
\end{table}

\section{How CPU cost was measured}\label{app:cpu}

\emph{CPU numbers come from one server, one library and one slide, with model loading excluded.} On a 12-core
CPU server (32\,GB RAM, no GPU) of a hospital-style container cluster, one binary built against numbat SDK 0.22.0
encoded tiles from one Yale-HER2 slide (read, JPEG decode, normalise, forward; f32). For each architecture we take
per-tile cost as the slope between two tile counts, so model loading and slide opening cancel; cost per slide is 400
times that slope.
ViT-B/16 was timed with our ViT-B/16 weights (Phikon's architecture) and ViT-g/14 with H-optimus-0. A build optimised
for that server encodes 400 tiles in 4.70--4.85\,s.

\end{document}